\documentclass[final,3p,times]{elsarticle}

\usepackage{graphicx}
\usepackage{amssymb}

\usepackage{lineno}
\usepackage{multirow}
\usepackage{pdflscape}
\usepackage{url}
\usepackage{adjustbox}
\usepackage{hyperref}
\usepackage{pdflscape}
\usepackage{afterpage}

\journal{MNRAS. Accepted 2021-12-15. Link: \url{https://doi.org/10.1093/mnras/stab3719}.}

\begin{document}

\begin{frontmatter}


\title{Recent formation and likely cometary activity of near-Earth asteroid pair 2019~PR2 -- 2019~QR6}


\author[ondrejov_address]{Petr Fatka\corref{cor1}}
\author[lowell_address]{Nicholas A. Moskovitz}
\author[ondrejov_address]{Petr Pravec}
\author[esa_neo_address,inaf_address]{Marco Micheli}
\author[arecibo_address]{Maxime Devog\`{e}le}
\author[sw_boulder]{Annika Gustafsson}
\author[lowell_address,ua_optics_address]{Jay Kueny}
\author[lowell_address]{Brian Skiff}
\author[ondrejov_address]{Peter Ku\v{s}nir\'{a}k}
\author[uni_arizona_address]{Eric Christensen}
\author[uni_texas_adress]{Judit Ries}
\author[uni_arizona_address]{Melissa Brucker}
\author[uni_arizona_address]{Robert McMillan}
\author[naval_address]{Jeffrey Larsen}
\author[uni_arizona_address]{Ron Mastaler}
\author[uni_arizona_address]{Terry Bressi}

\address[ondrejov_address]{Astronomical Institute of the Czech Academy of Sciences, Fri\v{c}ova 298, Ond\v{r}ejov CZ-25165, Czech Republic}
\address[lowell_address]{Lowell Observatory, 1400 West Mars Hill Road, Flagstaff, Arizona, USA}
\address[esa_neo_address]{ESA NEO Coordination Centre, Largo Galileo Galilei, 1, 00044 Frascati (RM), Italy}
\address[inaf_address]{INAF - Osservatorio Astronomico di Roma, Via Frascati, 33, 00040 Monte Porzio Catone (RM), Italy}
\address[arecibo_address]{Arecibo Observatory, University of Central Florida, HC-3 Box 53995, Arecibo, PR 00612, USA}
\address[sw_boulder]{Southwest Research Institute, 1050 Walnut St. 300, Boulder, CO 80302, USA}
\address[ua_optics_address]{Wyant College of Optical Sciences, University of Arizona, 1630 E. University Blvd., Tucson, AZ 85721, USA}
\address[uni_arizona_address]{The University of Arizona, Lunar and Planetary Laboratory, 1629 E. University Blvd.,Tucson, AZ 85721, USA}
\address[uni_texas_adress]{Department of Astronomy, University of Texas at Austin, 1 University Station C1400, Austin, TX 78712-0259, USA}
\address[naval_address]{Physics Department, United States Naval Academy 572C Holloway Road, Annapolis MD 21402, USA}

\cortext[cor1]{E-mail: fatka@asu.cas.cz}

\begin{abstract}
Asteroid pairs are genetically related asteroids that recently separated ($<$few million years), but still reside on similar heliocentric orbits. A few hundred of these systems have been identified, primarily in the asteroid main-belt. Here we studied a newly discovered pair of near-Earth objects (NEOs): 2019~PR2 and 2019~QR6. Based on broadband photometry, we found these asteroids to be spectrally similar to D-types, a type rare amongst NEOs. We recovered astrometric observations for both asteroids from the Catalina Sky Survey from 2005, which significantly improved their fitted orbits. With these refinements we ran backwards orbital integrations to study formation and evolutionary history. We found that neither a pure gravitational model nor a model with the Yarkovsky effect could explain their current orbits. We thus implemented two models of comet-like non-gravitational forces based on water or CO sublimation. The first model assumed quasi-continuous, comet-like activity after separation, which suggested a formation time of the asteroid pair $300^{+120}_{-70}$ years ago. The second model assumed short-term activity for up to one heliocentric orbit ($\sim$13.9 years) after separation, which suggested that the pair formed 272$\pm$7 years ago. Image stacks showed no activity for 2019~PR2 during its last perihelion passage. These results strongly argue for a common origin that makes these objects the youngest asteroid pair known to date. Questions remain regarding whether these objects derived from a parent comet or asteroid, and how activity may have evolved since their separation.

\end{abstract}


\end{frontmatter}


\section{Introduction}
Asteroids that experienced recent (less than a few Myr ago) break-ups have been known for over a decade with more than 200 systems of two (referred to as asteroid pairs) or more members (referred to as asteroid clusters or minifamilies). The vast majority of these systems are located in the Main Belt of asteroids between the orbits of Mars and Jupiter. The probable formation mechanism for the majority of these systems is rotational fission\footnote{Among other proposed formation mechanisms that might act in a fraction of known asteroid pairs, \cite{Vokrouhlicky17} suggested that the Datura family may be the result of a collision of a small projectile onto a nearly critically rotating parent body.} \citep{Scheeres07, Pravec10, Pravec18, Pravec19}. A rubble pile asteroid can be spun up by the Yarkovsky-O'Keefe-Radzievskii-Paddack (YORP) effect to its critical rotation rate, which leads to an escape of  material. Details of the mass loss process, e.g., whether a single large chunk of the asteroid is suddenly separated \citep[e.g.,][]{Scheeres07} or small pieces escape over some period of time \citep[e.g.,][]{Walsh08}, are still unknown. According to \cite{Pravec10} \textit{Supplementary Information}, after a short period of time (typically $< 1$ yr) the newly formed secondary asteroid may escape from the gravitational bound of its parent asteroid via spin-orbit interaction\footnote{Other possible more complicated scenarios have been discussed in \cite{Scheeres07, Jacobson11}}. The escape happens at a relative velocity comparable to the escape velocity from the larger body of the system (typically a few m/s for known pairs), which means that the separated asteroids have extremely similar heliocentric orbits for many years following the formation. The main belt of asteroids is for the most part dynamically stable, therefore it has been an ideal place to search for asteroid systems that recently formed via rotational fission \citep{Nesvorny06a, Nesvorny06b, Vokrouhlicky08, Pravec09}.

On the other hand, finding young, dynamical systems of asteroids or cometary nuclei, including those formed by rotational fission, in proximity to Earth’s orbit has proven to be much more challenging \citep[e.g.,][]{Fu05,Schunova12}. This is because the population of near-Earth asteroids is much smaller than the main belt asteroid population and the short coherence time of orbits in near-Earth space limits the search for dynamically associated objects to only the most recent breakups. Near-Earth objects (3200) Phaethon, (155140) 2005~UD and (225416) 1999~YC are an exception and their common origin is very likely \citep[e.g.,][]{Ohtsuka06, Jewitt06, Ohtsuka08, Hanus18} despite its difficult verification by $n$-body integrations due to the relatively long time since their separation and the general chaoticity of NEO orbits. However, certain ancillary evidence supports the common origin scenario for this cluster, such as, highly similar spectra suggesting rare B-type taxonomic classifications for the two largest members of the cluster and the almost identical polarization properties of their surfaces \citep{Devogele20}. \cite{Moskovitz19} identified two probable near-Earth asteroid pairs 2015~EE7 – 2015~FP124 (both members within the spectral taxonomic S-complex) and 2017~SN16 – 2018~RY7 (both rare V-types) and they constrained their separation ages to $<10^4$ yr.

In August 2019 two near-Earth objects (NEOs) were discovered on very similar heliocentric orbits (Table \ref{tab.orbits}): 2019 PR2 by the Pan-STARRS survey in Hawaii \citep{Chambers16} and 2019 QR6 by the Catalina Sky Survey (CSS) in Arizona \citep{Christensen19}, hereinafter referred to as "PR2" and "QR6", respectively. Based on their absolute magnitudes ($H$), these objects have a roughly 2:1 size ratio, with the primary about 1 km in diameter. These objects immediately stood out from typical NEOs due to their high eccentricity ($>0.79$) and, based on an early orbit solution, a semi-major axis that seemed to be very close to that of Jupiter. Future refinement of the orbits pushed their semi-major axes out beyond Jupiter. By September 2019 it was clear that these were in fact two separate objects that appeared to be flying in formation, only a degree apart on the sky. Such proximity, both in the space of orbital elements as well as physically (they were roughly 1 million km, or 0.007 AU separated in late September), was a strong indication that these objects were indeed genetically related and potentially formed in a recent separation event.

\begin{table}[h!]
\small
\caption{Vital stats for 2019 PR2 and 2019 QR6: absolute magnitude (H), and orbital elements semi-major axis ($a$), eccentricity ($e$), inclination ($i$), longitude of ascending node ($\Omega$), argument of perihelion ($\omega$), and mean anomaly ($M$) from our updated orbit solution including pre-covered astrometry data from 2005 (see Section \ref{sec.observations}). Orbital elements are derived for epoch JDT=2458800.5 (2019-Nov-13.0 TT; J2000)  and precision extends to the last significant digit in each case.}
\begin{center}
\begin{tabular}{llllllll}
Object	    & H [mag]		& $a$ [au]   & $e$ [~]     & $i$ [$^\circ$] & $\Omega$ [$^\circ$]  & $\omega$ [$^\circ$]  & $M$ [$^\circ$]\\
\hline
2019 PR2    & 18.8          & 5.77194240 & 0.79768236 & 10.989905 & 349.03448 & 57.09584 & 1.3151803\\
2019 QR6    & 20.0          & 5.77262648 & 0.79772000 & 10.971799 & 348.99833 & 57.12940 & 1.301104\\
\hline
\end{tabular}
\end{center}
\label{tab.orbits}
\end{table}%

Asteroids PR2 and QR6 are both near-Earth asteroids (NEA) with perihelion distance $q=1.16$ au, and are classified as Amor-type asteroids. Due to their semi-major axis $a = 5.77$ au and relatively large eccentricity $e = 0.8$, they spend most of the time beyond the orbit of Jupiter. Tisserand's invariant calculated with respect to Jupiter $T_{Jup} = 2.149$ puts these objects into the Jupiter-family comet group\footnote{Jupiter-family comets have $T_{Jup}$ between 2 and 3, while most asteroids have $T_{Jup} > 3$.}, however, many exceptions to this classification are known. 

Orbital similarity can be quantified with the Drummond $D$ criterion \citep{Drummond00}. For  PR2 and QR6 $D=1.3\times10^{-4}$, which is almost $30\times$ smaller than the exceptional cases 2015~EE7 – 2015~FP124 and 2017~SN16 – 2018~RY7 identified by \cite{Moskovitz19}. This makes them the two most similar orbits in the known near-Earth asteroid population of more than $25\,000$ objects (as of June 2021) according to analysis of Lowell Observatory's $astorb$\footnote{Accessible at \url{https://asteroid.lowell.edu}.} database of orbital elements. 
Motivated by the intriguing properties of this unusual pair of asteroids, we carried out characterization observations from October 2019 through March 2020 to measure broad band colors, rotational lightcurves, and astrometry for orbital refinement (Section \ref{sec.observations}). This collective set of observations, in addition to the retrieval of faint detections of both objects in CSS images 14 years prior to their discovery (during their previous perihelion passage), provided valuable insight into the origins of this pair. With a refined orbit solution based on 14 years of observations we were able to conduct deterministic backwards orbit integrations for several hundred years (Section \ref{sec.orbital_integ}). Consideration of several orbital evolution scenarios that included treatment of the Yarkovsky effect and non-gravitational accelerations due to comet-like out-gassing suggested a very recent separation of these objects. The cumulative results of these dynamical and observational analyses provided a means to interpret the origin and formation of the system (Section \ref{sec.formation}). A summary of our findings is provided at the end of this manuscript (Section \ref{sec.summary}).

\section{Observations and Data Analysis}
\label{sec.observations}

Images of PR2 and QR6 were obtained across several nights in October 2019 at the 4.3 m Lowell Discovery Telescope (LDT) in Happy Jack, Arizona, at the 2.3 m Bok telescope at Kitt Peak, and at the 2.1 m Otto Struve telescope at McDonald Observatory (Table \ref{tab.photo_obs}). Additional astrometric monitoring continued at LDT with observations in February and March when the two objects were too faint to be detected as part of regular operations by ongoing surveys (e.g. Pan-STARRS, CSS). Those astrometric measurements were reported to and are accessible at the Minor Planet Center.

\begin{table}[h!]
\footnotesize
\caption{Summary of observations for asteroids 2019~PR2 and 2019~QR6. See text for additional details and definitions.}
\label{tab.photo_obs}
\begin{tabular}{ccccccc}
Observatory     & Telescope     & Instrument        & UTC Date      & Filter    & V mag  & Exposures          \\ \hline
\multicolumn{1}{l}{\textbf{2019 PR2}} & & & & & \\
Lowell          & 4.3 m LDT     & LMI               & 2019-10-03    & SDSS griz & 18.7   & $17 \times  15$ s \\
                &               &                   & 2019-10-18    & SDSS griz & 18.8   & $36 \times 20$ s \\
Steward         & 2.3 m Bok     & SCC               & 2019-10-05    & SCC       & 18.7   & $348 \times 30$ s \\
                &               &                   & 2019-10-08    & SCC       & 18.7   & $295 \times 30$ s \\
McDonald        & 2.1 m Otto Struve & CQUEAN        & 2019-10-26    & SDSS i    & 18.9   & $132 \times 75$ s\\
                &               &                   & 2019-10-27    & SDSS i    & 18.9   & $136 \times 80$ s \\
                &               &                   & 2019-10-28    & SDSS i    & 18.9   & $158 \times 80$ s \\
\multicolumn{1}{l}{\textbf{2019 QR6}} & & & & & \\
Lowell          & 4.3 m LDT     & LMI               & 2019-10-03    & SDSS griz & 19.9   & $15 \times 20$ s \\
                &               &                   & 2019-10-18    & SDSS griz & 20.0   & $32 \times 30$ s \\
Steward         & 2.3 m Bok     & SCC               & 2019-10-06    & SCC         & 19.9   & $351 \times 30$ s \\
                &               &                   & 2019-10-07    & SCC         & 19.9   & $350 \times 30$ s
\end{tabular}
\end{table}

At LDT the observations were performed with the Large Monolithic Imager (LMI), which delivers a square 12.3' field-of-view sampled at 0.12~"/pixel. Images were obtained with 3x3 binning, tracking at sidereal rates, and using SDSS griz filters. The observations at LDT focused on measuring broad-band colors to determine  spectro-photometric properties. Image sequences with LMI involved sequentially cycling through the g, r, i, and z filters so as to provide a means to correct the colors for lightcurve variations in brightness.

The Spacewatch Cassegrain Camera (SCC) was employed at Steward Observatory's 2.3 m Bok telescope. SCC delivers at square 5' field-of-view sampled at 0.15~"/pixel. Images were obtained tracking at sidereal rates with the SCC filter, which provides broad transmission from 515-950 nm, roughly encompassing the Johnson-Cousins V, R and I bands. The Bok data were obtained to measure rotational lightcurves.

The CQUEAN (Camera for QUasars in EArly uNiverse) instrument \citep{Park12} was employed at McDonald Observatory's 2.1 m Otto Struve telescope. CQUEAN images a square 4.7' field at 0.276~"/pixel. Images were obtained using an SDSS i filter with the telescope tracking at sidereal rates. The data were obtained to measure the lightcurve of PR2.

Reduction of all images involved standard flat field and bias correction followed by measuring  photometry with the \texttt{Photometry Pipeline} \citep{Mommert17}. The Photometry Pipeline automatically registers images based on the Gaia catalog \citep{Gaia18}, extracts point source photometry using SourceExtractor \citep{Bertin96}, obtains photometric zero points for each image based on matching field stars to the PanSTARRS catalog \citep{Flewelling20}, and finally extracts calibrated asteroid photometry by querying the JPL Horizons system \citep{Giorgini97} for the position of the asteroid in each field. In general, 20-30 PanSTARRS field stars with solar-like colors (i.e., within 0.2 magnitudes of the Sun's g-r and r-i color) were used in every image to compute photometric zero points.

\subsection{Astrometry and orbital properties}
Using the observed arc for the objects during the 2019 apparition, it was possible to determine the ephemeris of both objects during their previous apparition, $\sim 14$ years earlier, with an estimated error of a few arcminutes. This level of uncertainty was sufficiently small to enable a search for  additional observations in image archives.

We performed a search for matching fields in the archive of the Catalina Sky Survey, which was fully operational in 2005. The search returned suitable fields exposed with the 1.5 m telescope on Mt. Lemmon, Arizona, USA, across three nights in October 2005. At this time, the two objects were expected to have reached an approximate magnitude $V \sim 21$, within reach of the telescope's capabilities. We carefully inspected the images from each night, checking the $3\sigma$ region around the line of variations of each target. Candidate moving sources were found on all three nights, and for both objects. 

Accurate astrometry was extracted for each of the nights, using a fourth order astrometric solution based on the Gaia DR2 stellar catalog. For PR2, the brighter member, we extracted multiple astrometric positions per night, while for QR6 a single stack of images from the first and last night produced sufficient signal to extract usable astrometry (Table \ref{tab.astrometry}).

The measurements obtained from these images have astrometric uncertainties of the order of $0.5''$. The candidate sources for each of the two objects were internally compatible with one other, and with the overall orbit fits, showing residuals comparable or better than the error bars, conclusively proving the correctness of the identification. 

When included in the overall orbital solution, these measurements improve the orbit accuracy by about 3 orders of magnitude, making the rest of the computations in this paper much more robust. Our updated orbit solution is shown in Table \ref{tab.orbits}.

\begin{table}[ht]
\centering
\caption{Astrometric measurements of the two objects obtained from
images in the archive of the Catalina Sky Survey. All exposures were
obtained with the Mt. Lemmon telescope (MPC code G96). The
measurements for 2019~PR2 correspond to individual frames, while the
first and last of those for 2019~QR6 correspond to stacks of all
images on those two nights.}
\begin{tabular}{lllll}
Object & Date (UTC) & $\alpha$ (hh:mm:ss.sss) & $\delta$ (dd mm ss.ss)
& Unc. ($''$) \\
\hline
2019 PR2 & 2005-10-27.102807 & 21:03:27.809 & -13 38 01.70 & 0.4 \\
2019 PR2 & 2005-10-27.108354 & 21:03:28.146 & -13 37 55.21 & 0.7 \\
2019 PR2 & 2005-10-30.096784 & 21:06:57.234 & -12 35 32.09 & 0.3 \\
2019 PR2 & 2005-10-30.102091 & 21:06:57.656 & -12 35 25.04 & 0.4 \\
2019 PR2 & 2005-10-30.107407 & 21:06:58.014 & -12 35 17.78 & 0.2 \\
2019 PR2 & 2005-10-30.112703 & 21:06:58.357 & -12 35 11.20 & 0.2 \\
2019 PR2 & 2005-10-31.080009 & 21:08:12.049 & -12 14 28.22 & 0.3 \\
2019 PR2 & 2005-10-31.086615 & 21:08:12.588 & -12 14 19.37 & 0.3 \\
\hline
2019 QR6 & 2005-10-27.104703 & 21:04:48.169 & -13 17 11.32 & 0.5 \\
2019 QR6 & 2005-10-30.097321 & 21:08:16.179 & -12 13 55.83 & 0.4 \\
2019 QR6 & 2005-10-30.113242 & 21:08:17.310 & -12 13 35.50 & 0.4 \\
2019 QR6 & 2005-10-31.083859 & 21:09:30.868 & -11 52 32.18 & 0.5 \\
\hline
\end{tabular}
\label{tab.astrometry}
\end{table}



\subsection{Broadband colors}

\begin{figure}[h!]
\centering
  \includegraphics[width=1.0\textwidth]{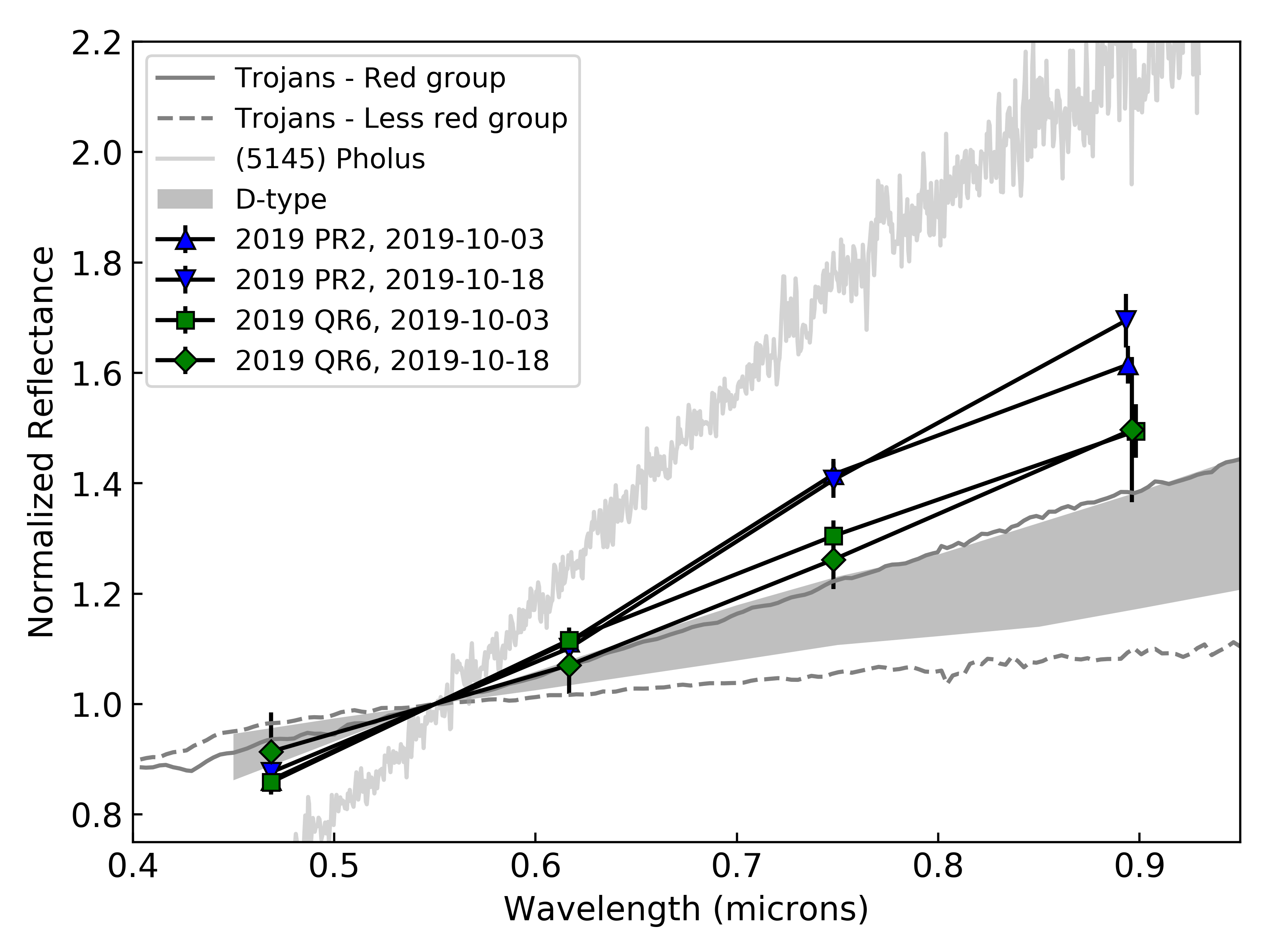}
  \caption{LDT {\it griz} spectro-photometry of PR2 (blue) and QR6 (green) from two different nights in October 2019. The inter-night agreement for each object is excellent. Both objects are best fit with a $D$-type taxonomic classification (grey region), though both are redder than the average $D$-type, as well as the red (solid grey) and less red (dashed grey) Trojan groups \citep{Emery09}. One of the reddest objects in the outer Solar System, Centaur (5145) Pholus \citep{Fornasier09}, is shown for context. The $z$-band points have been horizontally offset by small amounts to more clearly see the error bars.}
  \label{fig.colors}
\end{figure}

Broadband photometry in the SDSS filter set was obtained to determine the spectro-photometric reflectance of each body. In general, asteroid pairs that share a common origin have similar reflectance properties \citep{Moskovitz12,Duddy13,Polishook14,Wolters14,Moskovitz19, Pravec19}. Figure~\ref{fig.colors} shows the {\it griz} colors plotted as normalized, solar-corrected reflectance for both PR2 and QR6. Data were obtained on two different nights to validate results. The inter-night agreement in the measured colors is excellent. The weighted means and standard errors on the colors of PR2 are $g-r=0.71 \pm 0.03$, $r-i=0.37 \pm 0.02$, and $i-z=0.20 \pm 0.02$. The values for QR6 are $g-r=0.71 \pm 0.03$, $r-i=0.28 \pm 0.03$, and $i-z=0.18 \pm 0.04$.

\begin{figure}[h!]
\centering
  \includegraphics[width=1.0\textwidth]{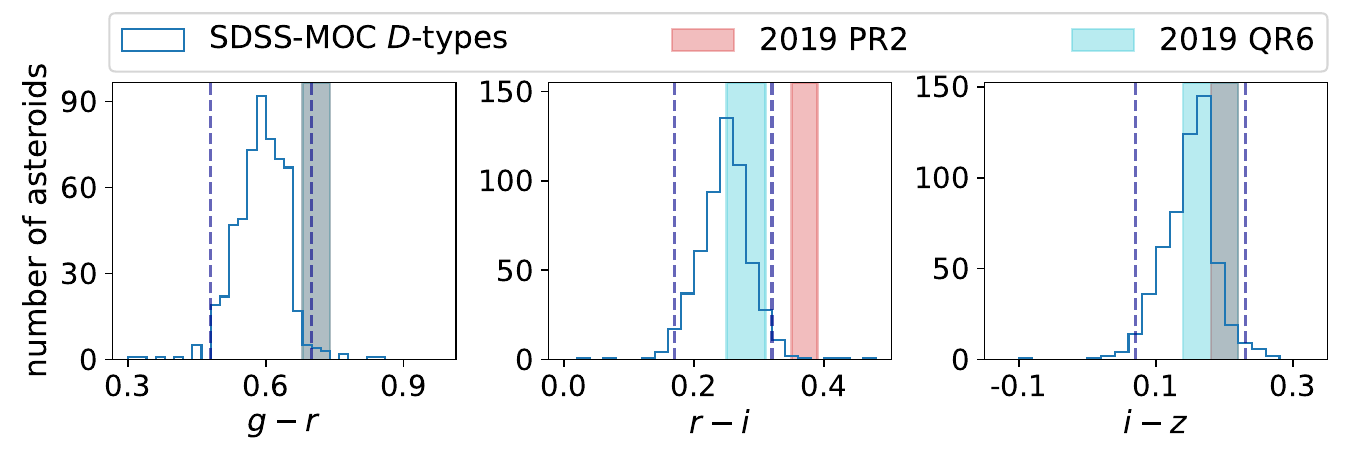}
  \caption{A comparison of color indexes of $D$-type asteroids \citep[classified by][]{Carvano10, Hasselmann11} from the the 4th release of the SDSS \textit{Moving Object Catalog} (blue histogram) and 2019~PR2 and 2019~QR6 (colored bars representing derived color indexes with their $1\sigma$ uncertainty). Dashed vertical lines represent the 2.275th and the 97.725th percentile of the SDSS distribution of $D-$type asteroids (simulating 2$\sigma$ deviation).}
  \label{fig.colors_compar_sdss}
\end{figure}

To assign a coarse taxonomic type to these objects, we computed a simple RMS residual between our data points and each of the Bus-DeMeo taxonomic classes \citep{DeMeo09}. In all four cases ---two each for PR2 and QR6--- we find that the data are best fit by a $D$-type classification. However, these objects are redder than the average $D$-type and lie outside of the 1-sigma envelope that defines the $D$ class (Fig. \ref{fig.colors}). These spectral properties are more often seen in populations in the middle and outer Solar System. While slightly redder than the Jovian Trojan Red Group \citep{Emery09}, PR2 and QR6 have shallower spectral slopes relative to very red objects like the Centaur (5145) Pholus. To demonstrate this further we compare the colors of our pair to that of $D$-types measured by SDSS and classified as such by \citet{Carvano10}. Figure \ref{fig.colors_compar_sdss} shows that PR2 and QR6 generally have colors about 1 to 2-sigma higher than the average SDSS $D$-type.

Our spectro-photometric data further show that PR2 has a significantly redder slope than QR6. Weighed averages of the slopes for PR2 and QR6 are $19.0 \pm 1.6~\%/0.1\mu$m~and $14.2 \pm 2.3~\%/0.1\mu$m~respectively. This difference is primarily attributed to the unusual $r-i$ color of PR2, which is significantly larger than that of QR6 and about 3-sigma off the mean $r-i$ color for SDSS $D$-types (Figure \ref{fig.colors_compar_sdss}). A number of factors may contribute to the difference in color between PR2 and QR6 including differences in observing geometry, object size, surface evolution, composition, and/or grain size. Each of these possibilities is discussed in Section \ref{sec.formation}.

\subsection{Rotational lightcurves}
\label{sec.rotations}

Asteroid PR2 was observed during five nights in its 2019 apparition from two observatories, while asteroid QR6 was observed for two consecutive nights 
from a single station (Table \ref{tab.photo_obs}). All Bok data taken in the SCC filter were calibrated to the SDSS $r$ filter, while the McDonald data were taken and calibrated in the SDSS $i$ band.  We processed and photometrically reduced all the nights using the Photometry Pipeline and the 3 nights 2019-10-26 to 28 for PR2 also independently with our Ond\v{r}ejov Observatory photometric reduction software package \texttt{Aphot32}.

We analyzed the reduced photometric data using the Fourier series method \citep[see][and references therein]{Pravec96}.  As the method assumes constant lightcurve shape, we confined our analysis of the PR2 data to the three consecutive nights 2019-10-26, 27 and 28.  The first two nights 2019-10-05 and 08 were taken at a substantially different aspect (with the asteroid's position in the sky different by about $25^\circ$) and the lightcurve shape could be significantly different (especially at the fairly high solar phases around $50^\circ$) at that epoch than that observed on the three later nights.  A joint analysis of all the five nights would require to model also a shape of the asteroid for which we do not have enough data.  For the period analysis, we used the data reduced at Ond\v{r}ejov that showed a somewhat lower scatter than the data reduced with the Photometry Pipeline, but we checked that an analysis of the Photometry Pipeline data gave very similar results and led basically to the same conclusions.

\begin{figure}[h!]
\centering
  \includegraphics[width=0.75\textwidth]{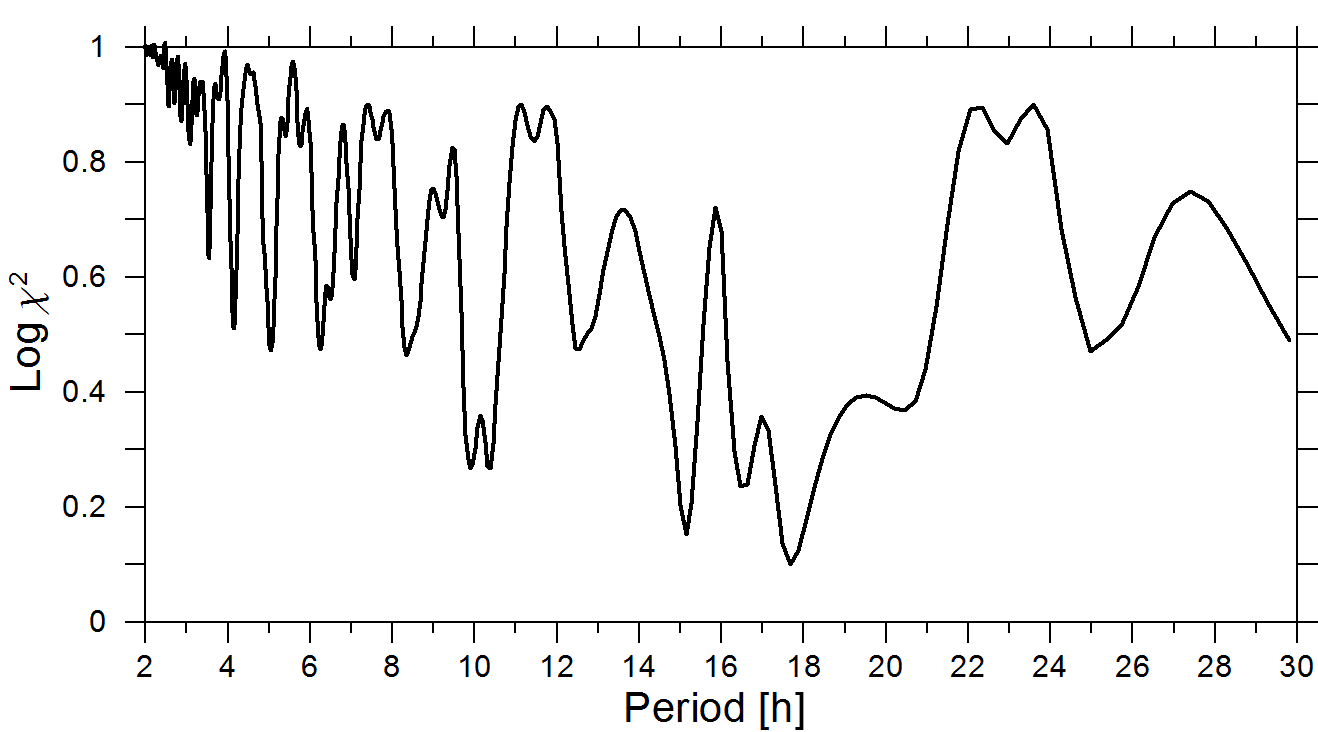}
  \caption{Noise spectrum for the 4th-order Fourier series fit to the 2019~PR2 lightcurve data 2019-10-26, 27 and 28.}
  \label{fig.pr2_rot_spec}
\end{figure}

The analysis of the three nights 2019-10-26 to 28 for PR2 revealed several possible periods in a range from 4.1 to 17.7~h.  In Fig. \ref{fig.pr2_rot_spec},
there are apparent several local minima in $\chi^2$ vs period for the 4th-order Fourier series fitted to the data, each of them corresponding to one possible period solution. We note that though the longer periods in the range 4.1--17.7~h fit formally better (have lower $\chi^2$), it may be only due to the limited length (coverage) of the observational data combined with possible flux (magnitude) calibration errors of (some of) the nightly sessions or a possible tumbling (see below).  So, the lower $\chi^2$ values for the longer periods may
be only formal and the shorter periods are possible as well.

\begin{figure}[h!]
\centering
  \includegraphics[width=0.75\textwidth]{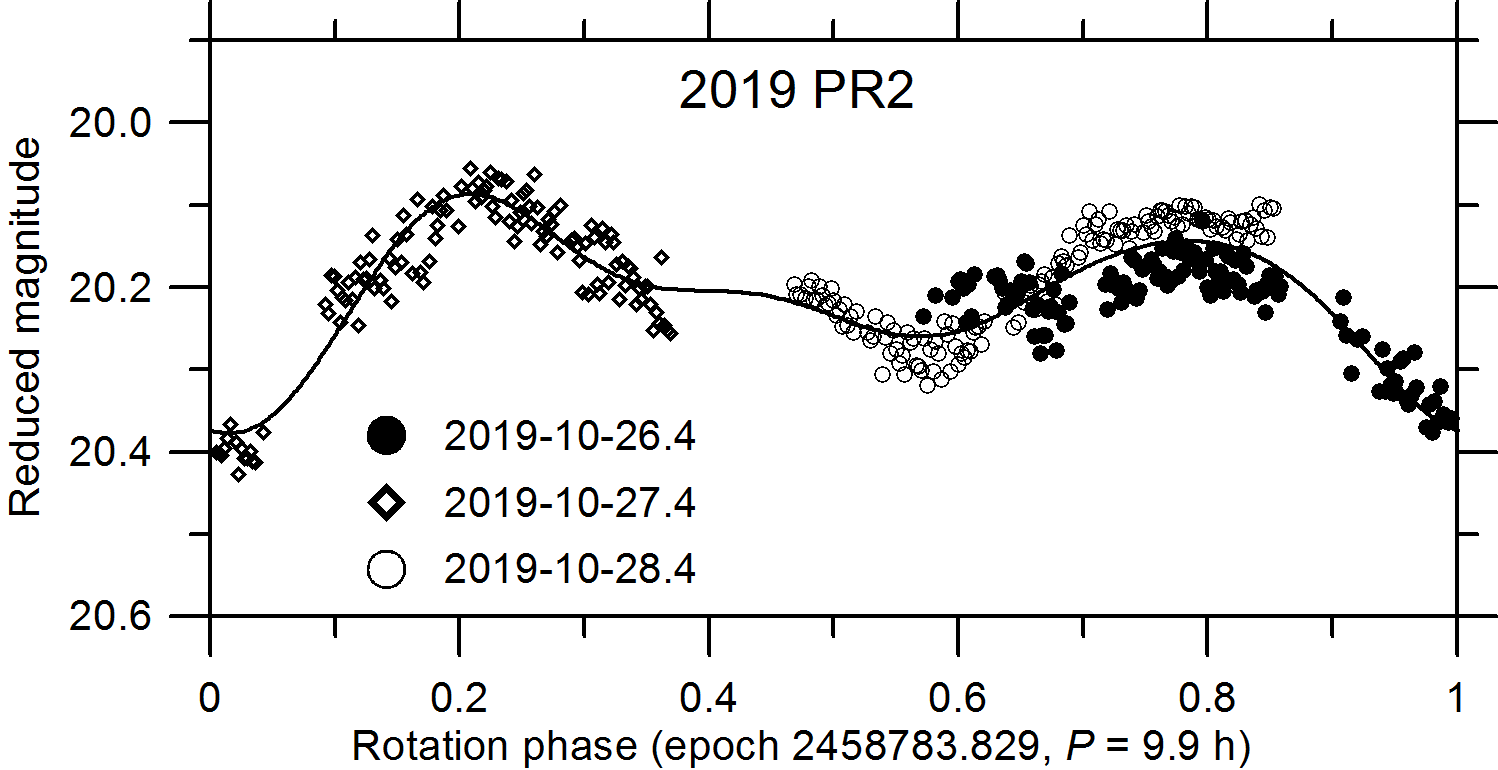}
  \caption{Composite lightcurve of 2019~PR2 for the most plausible rotation period of 9.9~h, for the epoch JD 2458783.829 (light-time corrected).  The SDSS $i$ magnitudes were reduced to unit geo- and heliocentric distances and to a consistent solar phase of $50^\circ$ using the $H$--$G$ phase relation with $G = 0.15$.  The curve is the best fit 4-th order Fourier series to the lightcurve data (points).}
  \label{fig.pr2_rot_pp}
\end{figure}

The most plausible of the several possible periods for PR2 appears to be 9.9 h, for which the lightcurve shape is most regular (bimodal, predominated by the 2nd harmonic), see Fig.~\ref{fig.pr2_rot_pp}. For the shorter periods in the range 4.1--8.4~h the composite lightcurves are monomodal (i.e., predominated by the 1st harmonic), while for the the longer periods 10.3--17.7~h the composite lightcurves have more complex shapes (with more extrema). However, such asteroid lightcurve shapes predominated by other than the 2nd harmonic are not impossible at the high solar phases ($53^\circ$) of the observations, so these periods are not ruled out.  More extensive observations taken around a future perihelion passage of the asteroid will be needed to resolve the period ambiguity.

A detailed check of how the data from the three nights fit together when folded with the candidate periods showed that they do not fold together perfectly for any of the possible periods.  In the composite lightcurve for the 9.9-h period (Fig.~\ref{fig.pr2_rot_pp}), the misfit is apparent as an offset of 0.06~mag between the data from the two nights 2019-10-26 and 28 at rotational phases 0.66--0.85.  Similar or larger misfits of the data from different nights are apparent also for the other possible periods in the range 4.1--17.7~h. The offsets or misfits between the data from different nights could be due to a non-principal axis rotation (so-called tumbling) of the asteroid (see discussion in Section~\ref{prim_rot_discus}), or they might be due to calibration errors of one or more of the observational nightly runs. Thorough and high-accuracy observations taken around a future perihelion passage of the asteroid will be needed to confirm the suspected tumbling of the asteroid PR2.

\begin{figure}[h!]
\centering
  \includegraphics[width=0.75\textwidth]{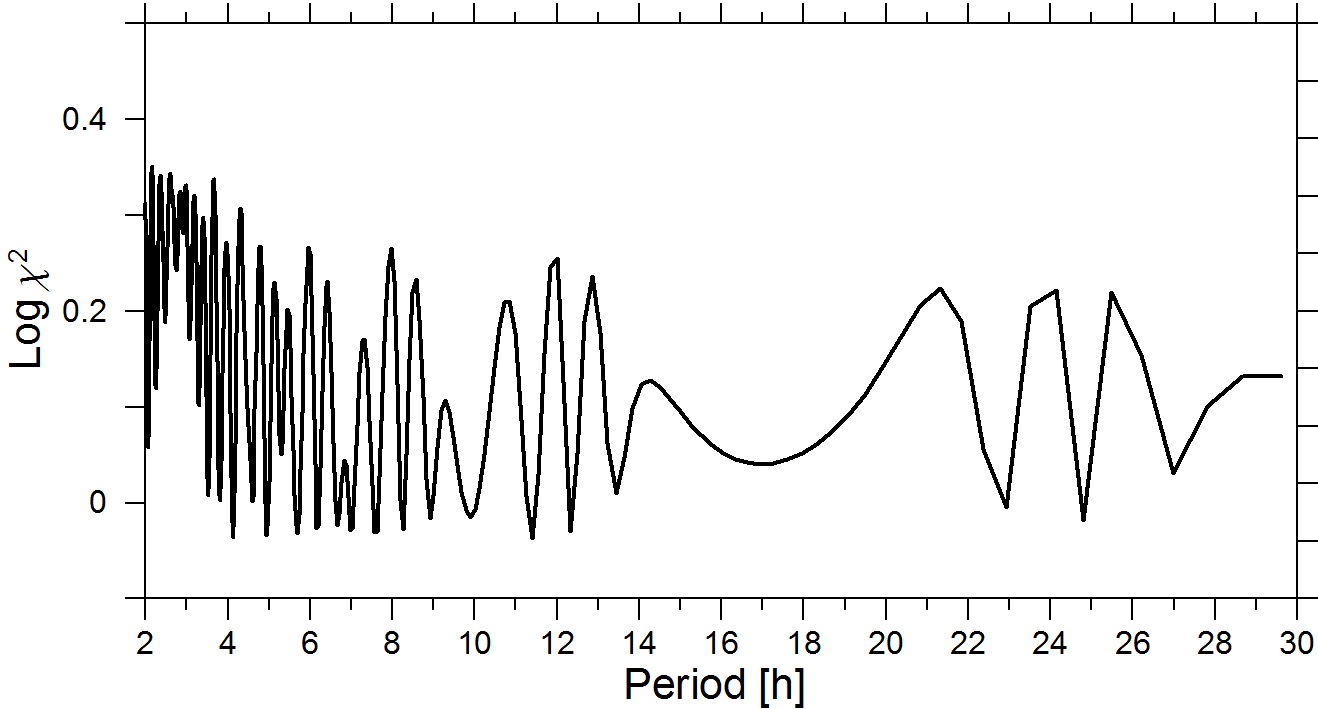}
  \caption{Noise spectrum for the 4th-order Fourier series fit to the 2019~QR6 lightcurve data 2019-10-06 and 07.}
  \label{fig.qr6_rot_spec}
\end{figure}

\begin{figure}[h!]
\centering
  \includegraphics[width=0.75\textwidth]{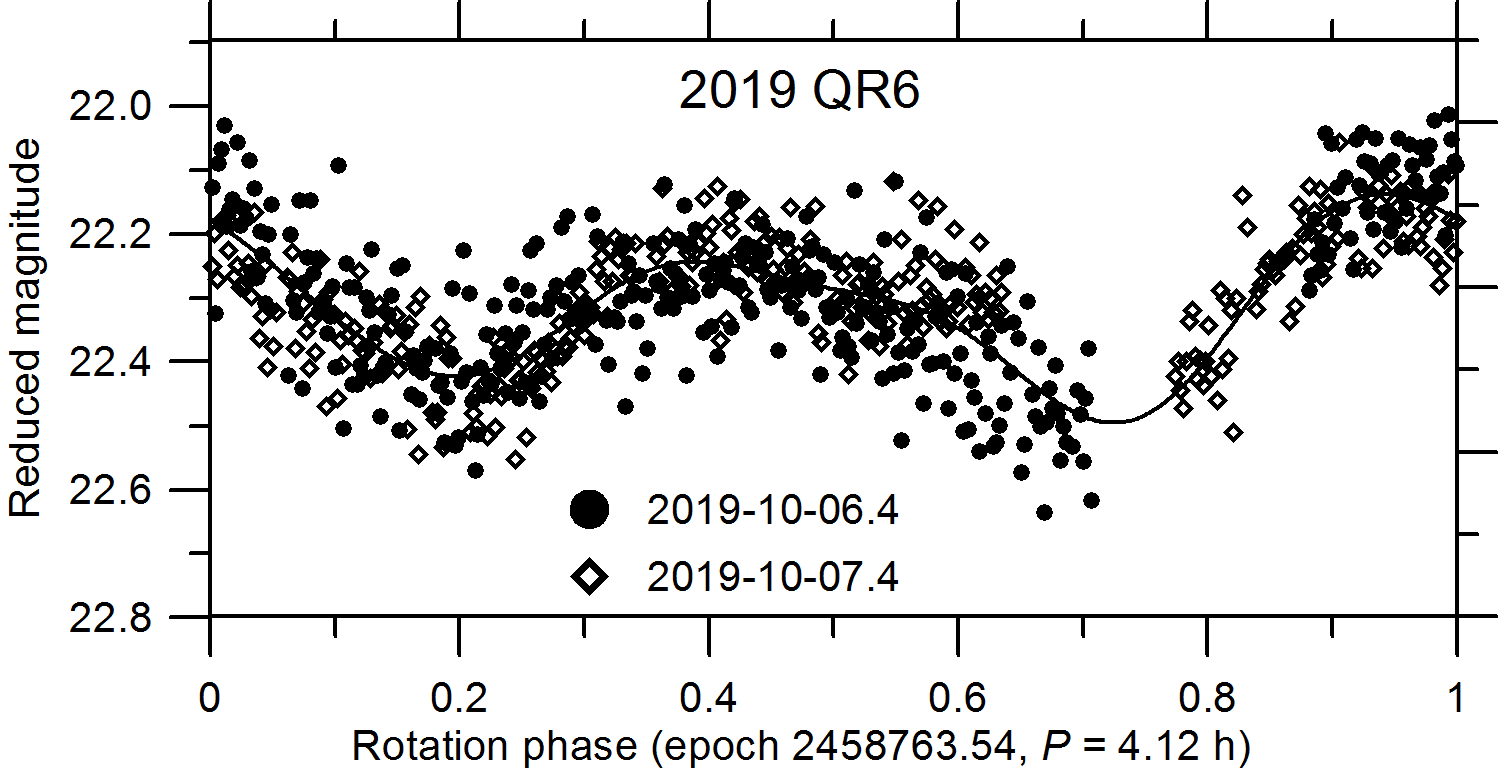}
  \caption{Composite lightcurve of 2019~QR6 for a candidate rotation period of 4.12~h, for the epoch JD 2458763.54 (light-time corrected).  The SDSS $r$ magnitudes were reduced to unit geo- and heliocentric distances and to a consistent solar phase of $50^\circ$ using the $H$--$G$ phase relation with $G = 0.15$.  The curve is the best fit 4-th order Fourier series to the lightcurve data (points).}
  \label{fig.qr6_rot_pp}
\end{figure}

An analysis of the two nights 2019-10-06 and 07 for QR6 tells a similar story, but its period ambiguity is even larger than for PR2 due to the QR6 lightcurve data being more limited (they are only two short nightly runs) and noisier. There are several possible periods for QR6 in a range from 3.5 to 5.0~h that have bimodal composite lightcurves, but a number of longer periods with more complex lightcurve shapes are also possible, with an upper limit of 13.5~h.  See the noise spectrum $\chi^2$ vs period in Fig. \ref{fig.qr6_rot_spec}.  (We note that though some of the local minima in $\chi^2$ go slightly below 1 ($\log \chi^2 < 0$), it is only a formal feature resulting from probably slightly overestimated errors of the data points.)
Given the limited amount of data, it is difficult to assess which of the several possible periods is the most likely or plausible; a composite lightcurve for one of them is shown in Fig.~\ref{fig.qr6_rot_pp}.  We also note that for this period analysis, we took the zero points of the magnitude scales of the two observing nights as free parameters in the fit.  For the best fit, we obtained their relative offset of 0.21~mag.  We consider it to be due to a magnitude calibration error in one or both nights rather than a real feature of the asteroid.  However, high accuracy (and more extensive) photometric observations will be needed for this asteroid around its future perihelion passage to reveal its spin state.

\subsection{Constraints on activity}
\label{sec.constrains-activity}

\begin{figure}[h!]
\centering
  \includegraphics[width=0.7\textwidth]{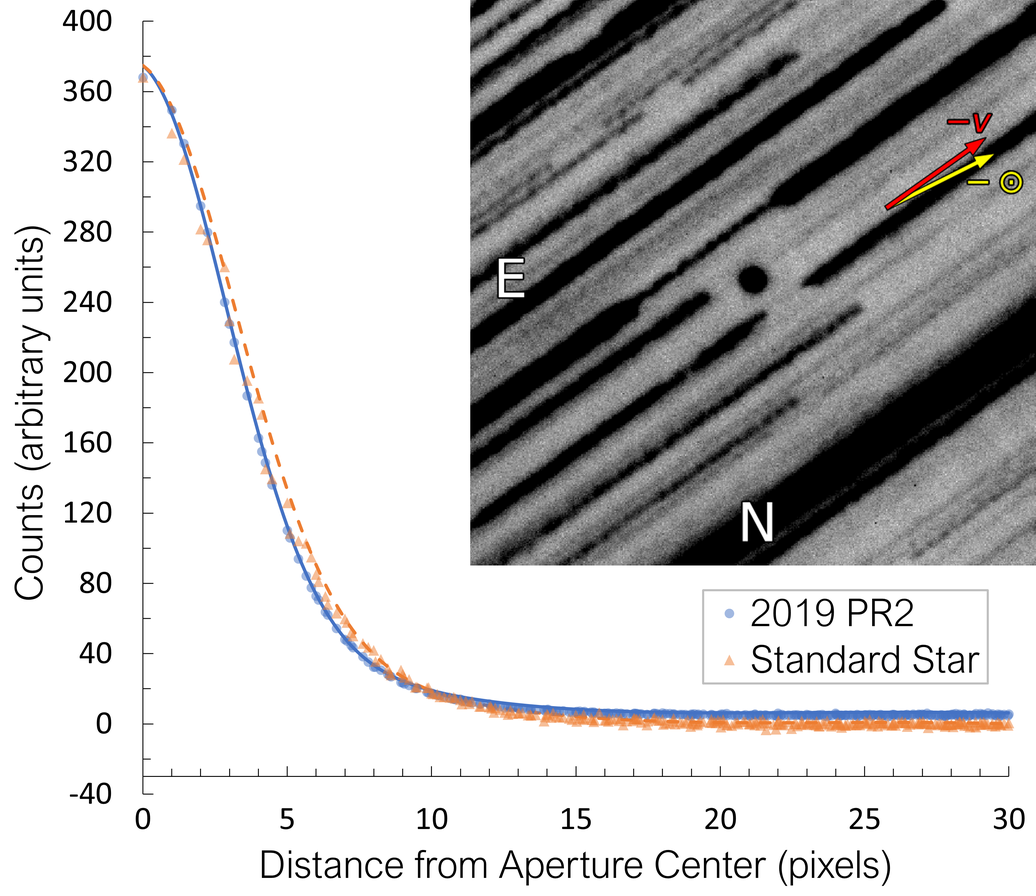}
  \caption{Radial profile of 2019 PR2 from a stack of 331 images taken on 2019-10-05 with the 2.3 m Bok, compared to a field star from those images. We see no significant differences in the PSFs, and the image stack (inset) shows no indication of faint tail or coma. The cardinal directions, anti-Sun vector, and anti-velocity vector are shown in the inset. Despite these images being taken near perihelion, PR2 shows no signs of recent activity.}
  \label{fig.psf}
\end{figure}

PR2 and QR6 are on eccentric, comet-like orbits and were imaged only a few weeks ahead of their perihelion passage on 2019-10-25. Based on precedent from asteroid (3200) Phaethon ---an eccentric, near-Earth asteroid, pair primary that displays low levels of activity around perihelion \citep{Jewitt10}--- we performed a search for indications of activity around PR2. We focused this analysis on the 2019-10-05 data from the 2.3 m Bok as these represent our deepest exposures when the object was not trailed in individual images. We median stacked the images at the non-sidereal rate of the asteroid and only included images where no field star was within $\sim 25$ pixels (3.6") of the asteroid center of brightness (Fig. \ref{fig.psf}). This provided 331 individual exposures of 30 s each for the image stack, yielding an effective exposure time of 9930 s. The profile of the asteroid in this deep stack was compared to that of a field star from a 6-image, sidereal-rate, mini-stack that represented the 6 images with the lowest average full-width-half-maximum (FWHM) of the point spread functions (PSFs) across all field stars. No significant difference was seen between the radial profile of the asteroid and the field star, indicating that our stacking procedure was robust and that PR2 did not host a low surface brightness coma. Visual inspection of the deep image stack further showed that PR2 did not have a dust tail or trail in the anti-solar or anti-velocity directions (Fig. \ref{fig.psf}).

To quantify the non-detection of a coma we took a median stack on the asteroid of all 331 images and measured the standard deviation of background counts in an annulus 13-20 pixels (roughly 2-3 times the FWHM) from asteroid center. That standard deviation was converted to a 3-sigma limiting magnitude using a conservative zero point per 30-second exposure of 29.7 (this is the zero point computed by the Photometry Pipeline and corresponds to a source that would produce 1 count per pixel per exposure, 99\% of our images have a zero point greater than this). This calculation suggests that PR2 showed no activity down to a 3-sigma limiting magnitude of 27.9 in SDSS $r'$.

\section{Orbital integrations}
\label{sec.orbital_integ}

To study the orbital history of PR2 and QR6 and to verify their common origin, we employed a backwards orbital integration method. Each asteroid was represented by a set of $10^4$ orbital clones to reflect initial orbital uncertainties. These clones were created using the probability distribution $p\left(\textbf{E}\right) \propto \exp \left( -\frac{1}{2} \Delta \textbf{E} \cdot \Sigma \cdot \Delta \textbf{E} \right)$, where \textbf{E} is a set of six orbital elements for a given clone, $\Delta \textbf{E} = \textbf{E} - \textbf{E}^{*}$ is the orbit difference with respect to the best-fit orbital values $\textbf{E}^{*}$ (Table \ref{tab.orbits}), and $\Sigma$ is the normal matrix of the orbital solution \citep{Milani10}. 

We used the \texttt{IAS15} integrator \citep{Rein15} from the \texttt{REBOUND} package \citep{Rein12}. \texttt{IAS15} is a high accuracy non-symplectic integrator with adaptive time steps, which correctly solves for close encounters between orbital clones (massless test particles) and massive bodies (e.g. planets). Since we are interested in the convergence of clones from each asteroid, we modified the original code to limit the maximum time-step to 1 day. In our integration we accounted for the gravitational attraction of the Sun, the 8 major planets, two dwarf planets Pluto and Ceres, and two large asteroids Vesta and Pallas.

Using the \texttt{Find\_Orb}\footnote{Available at \url{https://www.projectpluto.com/find_orb.htm}} software tool, we got the best fit orbital solutions for each asteroid using our full set of astrometry (Section \ref{sec.observations}). We tested orbital fits assuming a pure gravitational model, a model including the Yarkovsky effect \citep{Farnocchia13}, and the Marsden model \citep{Marsden73} describing accelerations caused by cometary activity. All initial orbits were estimated for a single epoch JDT = 2458800.5 (2019-11-13.0 TT), which is near the center of the most recent apparition. 
The best orbital fits (without any non-gravitational forces) are shown in Table \ref{tab.orbits}.

The chosen epoch JDT = 2458800.5 is also the start point of our backward integrations and the integration time $T$ is stated relative to this epoch. Results from individual integrations are discussed in the following subsections. 

In all cases we integrated the orbital clones backwards for $1\,000$ years to search for close (and preferably slow) encounters, which are expected for objects that share a common origin. Rotational fission is a widely accepted formation mechanism of most asteroid pairs and clusters and it is associated with separation velocities on an order of m/s \citep[e.g.,][]{Scheeres07,Vokrouhlicky08, Pravec19}.

\subsection{Pure gravitational model} 
\label{sec.pure_grav}

\begin{figure}[h!]
\centering
  \includegraphics[width=0.9\textwidth]{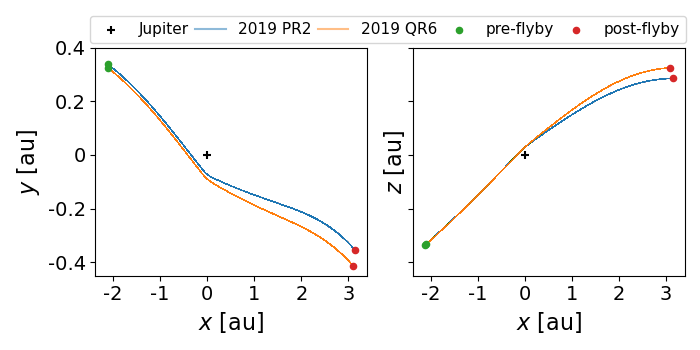}
  \caption{Trajectories of $10^4$ orbital clones of asteroid PR2 (blue line) and QR6 (orange line) during a close encounter with Jupiter (black plus sign) around $T\sim-406$ years in the Jovicentric coordinates. The time interval from initial position, marked by green points, ($T = -405.3$ yr) to final positions ($T = -407.3$ yr), marked by red points, is 2 years long. \textbf{The orientation of $x,y,z$ axes is identical with the \textit{International Celestial Reference Frame} (third revision).}} 
  \label{fig.jup_encounter}
\end{figure}

For both asteroids we derived orbits excluding non-gravitational forces (Table \ref{tab.orbits}), populated the uncertainty regions with clones, and performed their backward orbital integration. We did not find any reasonably close encounters between the clones of PR2 and of QR6 during the integration. The smallest relative distances between clones were $\sim80\,000$ km, which we consider too large given the precise initial orbits, integrator reliability, and the relatively short integration time. At $T\sim -406$ yr both sets of clones experienced a relatively close encounter with Jupiter with the closest approaches about $(11.51 \pm 0.03)\times 10^6$ and $(13.88 \pm 0.03) \times 10^6$ km from Jupiter's system barycenter for PR2 and QR6, respectively. We also tested the possibility of a close encounter with Jupiter's moons Himalia and Elara (orbiting at distances $\sim 11.5 \times 10^6$ km around Jupiter), but no encounters were found. We additionally found that the mean times of the closest approach for each set of clones differed by $\sim36$ hours, whereas the times of the closest approach to Jupiter for clones of the same asteroid happened within 1 hour for PR2 and within 2 hours for QR6. Prior to this Jovian encounter, the orbits of both clone sets started to differ from one another significantly while remaining as compact clouds (see Fig. \ref{fig.jup_encounter}). This suggests that the encounter with Jupiter was solved properly and that the orbits remained deterministic. It is extremely unlikely that two unrelated asteroids (with the same taxonomic classification) would experience an encounter with Jupiter that would alter their orbits in such way that they would appear on the very similar orbits we observe today. Therefore, we suspect that non-gravitational forces must play a role it the recent history of PR2 and QR6.


\subsection{Inclusion of the Yarkovsky effect}
\label{sec.yarkovsky_effect}
We tested the influence of the Yarkovsky effect on these asteroids' orbital history. We simulated the Yarkovsky effect as a transverse acceleration $\dot{a}_{\rm{Yarko}}$ acting on each clone, producing a secular change in the semi-major axis \citep{Farnocchia13}. Each clone was assigned with a random value of $\dot{a}_{\rm{Yarko}}$ from a range of possible values $\langle -\dot{a}_{\rm{max}}, \dot{a}_{\rm{max}}\rangle$, where $\dot{a}_{\rm{max}}$ is the maximum semi-major drift scaled to the asteroid's estimated size. The effective asteroid diameter was estimated using the formula $D = \frac{1329}{\sqrt{p_{\rm{V}}}} 10^{-H/5}$, where $p_{\rm{V}}$ is the geometric albedo \citep[e.g.,][]{Fowler92}. However, we do not know the geometric albedo for either of these asteroids, therefore we can only assume that its value is a few percent based on the associated $D$-type taxonomic classification \citep[e.g.,][]{Fitzsimmons94}. For completeness, we tested a wide range of values for $p_{\rm{V}}$ from 0.02 up to 0.5. Even for the highest values of $p_{\rm{V}}$ (meaning smallest asteroid sizes and strongest Yarkovsky effect), we did not find any close encounters between the clones of PR2 and QR6. The smallest recorded distances during slow encounters were smaller only by several thousand kilometers than the closest encounters at $\sim80\,000$ km recorded in the pure gravitational integration. None of the clones in any of these integrations (with Yarkovsky effect included) avoided the Jovian encounter seen in the pure gravitational integration, which put the two clone sets into significantly different orbits. This suggests that other non-gravitational forces must have been involved in the recent dynamical history of this asteroid pair.

\subsection{Post break-up activity models}
\label{sec.post_breakup_activity}
Since the inclusion of the Yarkovsky effect did not explain the formation of the asteroid pair, we turned to developing a new model that includes cometary-like non-gravitational forces acting on the bodies. For that, we used a model of symmetric non-gravitational accelerations, originally developed to describe motions of active comets \citep{Marsden73}. A radial component of the non-gravitational acceleration is expressed as $A_1 g(r)$, a transverse component as $A_2 g(r)$, and a component normal to the orbital plane as $A_3 g(r)$. Here $A_i$ are constants and 
\begin{equation}
\label{eq.gr-function}
    g(r) = \alpha \left( \frac{r}{r_0} \right)^{-m} \left[1 +  \left( \frac{r}{r_0} \right)^n \right]^{-k},
\end{equation}
where $r$ is the heliocentric distance and ($\alpha, r_0, m, n, k$) are coefficients, whose values are in Table \ref{tab.gr_coef} \citep{Krolikowska17, Yabushita91}.

\begin{table}
\caption{Coefficients of the $g(r)$-function for the water sublimation model \citep{Marsden73} and CO sublimation models \citep{Krolikowska17, Yabushita91}. We note that $\alpha$ is a normalization constant chosen in order that $g(1\rm{au}) = 1$.}
\label{tab.gr_coef}
\begin{tabular}{l|lllll}
            & $\alpha$ & $r_0$ & $m$   & $n$   & $k$     \\ \hline
water sublim. & 0.1113   & 2.808 & 2.15 & 5.093 & 4.6142 \\
CO sublim.  & 0.01003  & 10.0  & 2.0  & 3.0   & 2.6   
\end{tabular}
\end{table}

\begin{figure}[ht]
\centering
  \includegraphics[width=0.6\textwidth]{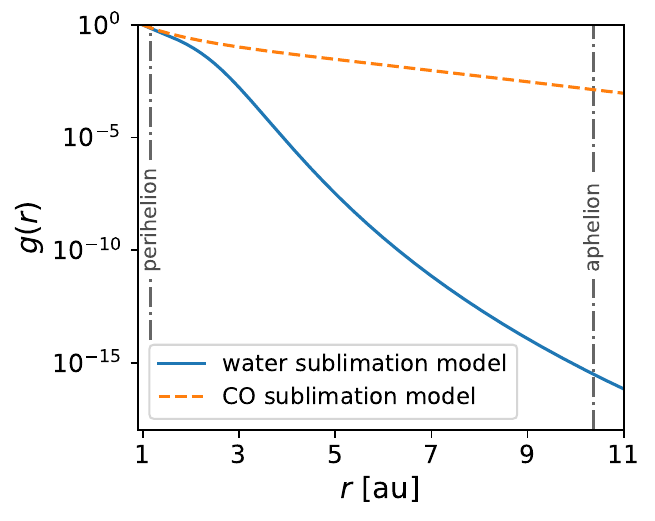}
  \caption{The non-gravitational acceleration function, $g(r)$, normalized at r=1 AU, with coefficients representing water and CO sublimation models. Perihelion ($q\approx1.2$ au) and aphelion ($Q\approx10.4$ au) distances of PR2 and QR6 current orbits are highlighted.}
  \label{fig.gr_func}
\end{figure}

\subsubsection{Continuous activity}
\label{sec.non-stop-model}
Using \texttt{Find\_Orb}, we derived PR2 and QR6 orbits for a water and a CO sublimation models (see Table \ref{tab.orbits_A123_appendix} in Appendix) with coefficients ($\alpha, r_0, m, n, k$), occurring in the $g(r)$ function (Eq. \ref{eq.gr-function}), listed in Table \ref{tab.gr_coef}. The formal best-fit $A_1, A_2, A_3$ coefficients, representing non-gravitational acceleration, are of order $10^{-10}$ au/d$^2$, and are comparable to their formal uncertainties (Table \ref{tab.A123_coef}). The only exception is the $A_3$ coefficient for PR2, which has a significance  at a $2\sigma$ level. Each clone was assigned with random $A_1, A_2, A_3$ values assuming normal distributions with means and standard deviations taken from the orbital fits. We tested whether such a weak (below the detection limit) non-gravitational acceleration could allow for close encounters between the clones to appear.

\begin{table}
\caption{Coefficients $A_1, A_2, A_3$ (with $1\sigma$ uncertainties) obtained from the best orbital fits for a model including free parameters representing non-gravitational acceleration \citep[following][]{Marsden73}.}
\label{tab.A123_coef}
\begin{tabular}{ccccc}
sublimation            & \multirow{2}{*}{asteroid} & $A_1 \times 10^{-10}$ & $A_2 \times 10^{-10}$ & $A_3 \times 10^{-10}$ \\
model                  &                           & {[}au/d$^2${]}        & {[}au/d$^2${]}        & {[}au/d$^2${]}        \\ \hline
\multirow{2}{*}{water} & 2019 PR2                  & $-1.7 \pm 2.9$        & $-0.1 \pm 2.1$        & $1.7 \pm 0.9$         \\
                       & 2019 QR6                  & $-3.1 \pm 3.5$        & $-0.3 \pm 2.7$        & $1.0 \pm 1.0$         \\ \hline
\multirow{2}{*}{CO}    & 2019 PR2                  & $-1.7 \pm 2.7 $       & $ 1.5 \pm 0.9$        & $2.0 \pm 1.0$         \\
                       & 2019 QR6                  & $-1.6 \pm 3.0$        & $1.3 \pm 1.1$         & $0.9 \pm 1.0$        
\end{tabular}
\end{table}

Even though the non-gravitational forces were technically applied over whole orbits, Figure \ref{fig.gr_func} shows that the resulting acceleration ($A_i g(r)$) is negligible at $r > 3$ au in the case of the water sublimation model, and it is only at 10\% of the maximum $A_i$ value at $r \sim 3$ au in the case of the CO sublimation model.

\begin{figure}[ht!]
\centering
  \includegraphics[width=1.1\textwidth]{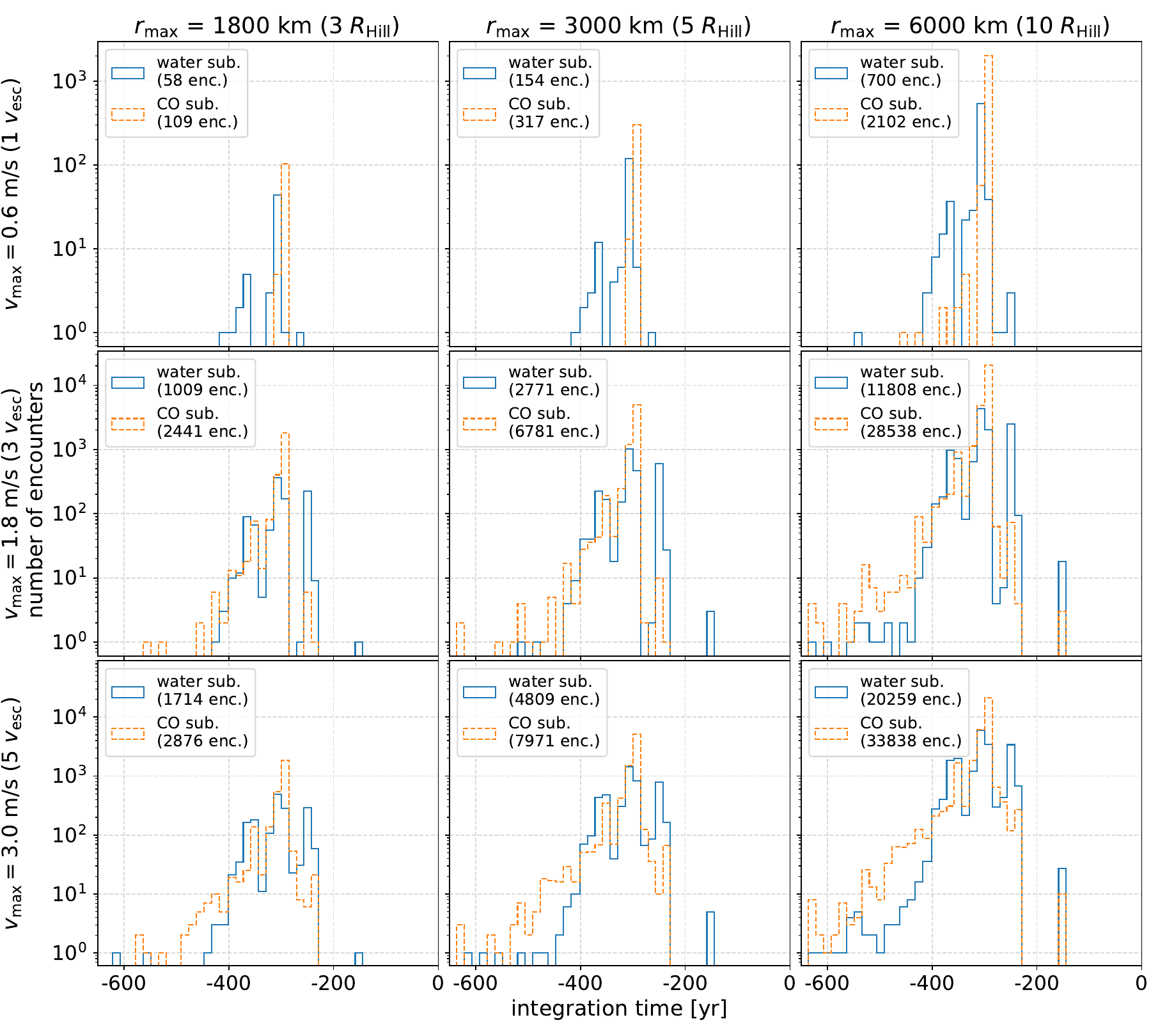}
  \caption{Histogram of close clone encounters for the water sublimation model (blue) and the CO sublimation model (orange). Each column represents a different maximum distance between clones for an encounter to be considered close and each row displays a different maximum relative velocity between clones for an encounter to be considered slow. The width of bins is chosen to be one full orbit of asteroid PR2 ($\sim 13.9$ years) around the Sun, which makes it the same width as in Fig. \ref{fig.encounters_hist}.}
  \label{fig.encounters_hist_non-stop}
\end{figure}

Resulting time distributions of clone encounters are shown in Figure \ref{fig.encounters_hist_non-stop}. For the strictest values of $v_{\rm{max}}$ and $r_{\rm{max}}$ (equal to $v_{\rm{esc}}$ and $3R_{\rm{Hill}}$, respectively, as expected from the rotational fission theory) the encounters happened at $T$ in range (-290, -310) yr with the median value -292 yr in the case of the CO sublimation model and a bimodial distribution is apparent for the water sublimation model with $T$ between (-260, -325) yr with the median value -309 yr for the first cluster of encounters and with $T$ between (-360, -401) yr with the median value -371 yr for the other one (see the top left panel of Fig. \ref{fig.encounters_hist_non-stop}).
For the relaxed limits of $v_{\rm{max}}$ and $r_{\rm{max}}$ the above times are still preferred, but many other encounters are detected, mainly towards lower $T$ (to the left in the histograms in Fig. \ref{fig.encounters_hist_non-stop}). The integrations with the CO sublimation model produces roughly twice as many encounters as the water sublimation model. This might be due to the overall higher values of the $g(r)$ function (see Fig. \ref{fig.gr_func}). Some clones are deflected from the close Jovian encounter due to the non-gravitational acceleration and we recorded several encounters at $T < -407$ yr, however they were only about 1.5\% of the recorded clone encounters for $v_{\rm{max}} = 5~ v_{\rm{esc}}$ and $< 1$\% in the remaining cases.

\begin{figure}[h!]
\centering
  \includegraphics[width=0.8\textwidth]{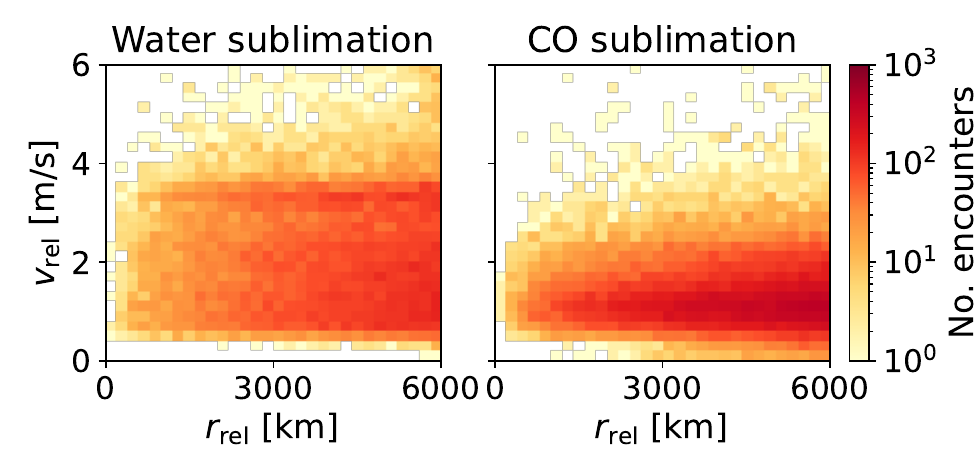}
  \caption{A heat-map visualizing number of clone encounters with given $v_{\rm{rel}}$ and $r_{\rm{rel}}$ values. $Left$ panel shows $v_{\rm{rel}}$ and $r_{\rm{rel}}$ distribution for the water sublimation model, $right$ panel shows the same distribution for the CO sublimation model.}
  \label{fig.enc_heatmap_non-stop}
\end{figure}

The distribution of $v_{\rm{rel}}$ and $r_{\rm{rel}}$ of clones during encounters is shown in Figure \ref{fig.enc_heatmap_non-stop}. Both models produced encounters with very low $v_{\rm{rel}}$ and $r_{\rm{rel}}$, but encounters in the CO sublimation model are much more compact and accumulated around $v_{\rm{rel}} \sim 1.2$ m/s (about $2 v_{\rm{esc}}$) despite being more numerous.

The results show that a weak continuous (i.e., with constant $g(r)$ function) non-gravitational acceleration, which would be caused by repeated periodic outgassing during each perihelion flyby, would be able to overcome the $\sim80\,000$ km gap between clones seen in the integrations using the model with the  Yarkovsky effect in the previous section. More than 98\% of all encounters happened at $T = -300^{+70}_{-120}$ yr regardless chosen $v_{\rm{max}}$ and $r_{\rm{max}}$ limits.

As we showed above, the clone orbit integrations with continous comet-like activity (i.e., with constant $g(r)$ function) could formally explain the current observed orbits of the asteroid pair with their separation time about 300~yr ago.  However, we do not consider this model to be a realistic scenario.  The reason is that it neither has been observed nor it appears plausible that asteroids would be active continuously with a constant $g(r)$ function for hundreds of years, and, as we demonstrated in Section~\ref{sec.constrains-activity}, we did not detect any signs of cometary activity for PR2 in its last apparition.

\subsubsection{One-orbit activity model}
\label{sec.one-orbit-model}
In this section we investigate the possibility of a short-lived phase of activity following rotational fission of a parent body. Our model here assumes that the two asteroids PR2 and QR6 were active from the time of their formation by rotational fission until the next aphelion passage (where the non-gravitational acceleration was negligible anyway due to the $g(r)$ function and $r \approx 10.4$ au). After the aphelion passage, the activity causing the non-gravitational forces stopped and the asteroid orbits evolved purely under the influence of gravity. This activity could last for up to one full orbit (if fission happened at aphelion), however, our model allows for the possibility that the break-up happened at any point of the heliocentric orbit (i.e., at a smaller heliocentric distance), in which case the activity would last for less than one full orbit.

However, our integrations must start at the current epoch, when the orbits are well constrained, and we must propagate the asteroids backwards into the past. Both asteroids completed 28 orbits around the Sun between the last observations in 2020 and the encounter with Jupiter at $T\sim-406$ years (Section \ref{sec.pure_grav}). We integrated sets of orbital clones back to each of the previous 28 aphelion passages without any non-gravitational forces. At each of these aphelia we switched on the non-gravitational forces, simulating the asteroids' activity, and integrated the system back in time for one more full orbit during which we searched for close encounters. Each of the 28 integrations was stopped after this single orbit of activity. Each clone was assigned with a random set of $A_1, A_2, A_3$ values from a uniform distribution in the range $\langle-A_{\rm{max}}, A_{\rm{max}}\rangle$, where $A_{\rm{max}}$ was a specified limit. For each of the 28 integrations, we tested several values of $A_{\rm{max}}$ ranging from $10^{-10}$ au/d$^2$ to $3.5\times10^{-8}$ au/d$^2$. We performed a detail study for three chosen values of $A_{\rm{max}}$, namely $10^{-8}$ au/d$^2$ as it is the lowest value for which close encounters appear, $2\times10^{-8}$ au/d$^2$ as it provides a reasonable number of detected encounters for a statistical analysis for the water sublimation model, and $2.5\times10^{-8}$ au/d$^2$ as it is the lowest value allowing for encounters with the CO sublimation model. More encounters can be obtained with higher values of $A_{\rm{max}}$ (see Fig. \ref{fig.encounters_hist_350_appendix} in Appendix), but we conservatively limited the detailed studies to just these three values. We note that larger values of $A_{\rm{max}}$ are possible, but results differ mainly in the number of detected encounters.

All of the tested values of $A_{\rm{max}}$ are smaller by at least two orders of magnitude than the mean value of non-zero $A_i$ parameters in the \textit{Catalogue of Cometary Orbits and their Dynamical Evolution}\footnote{Accessible at \texttt{https://pad2.astro.amu.edu.pl/comets/}}, which contains mainly long-period comets. The $A_i$ values are expected to be higher for pristine and preserved comets during their few returns, therefore, the notably lower values of $A_{\rm{max}}$ required in our integrations meets our expectation.

\begin{figure}[h!]
\centering
  \includegraphics[width=1.1\textwidth]{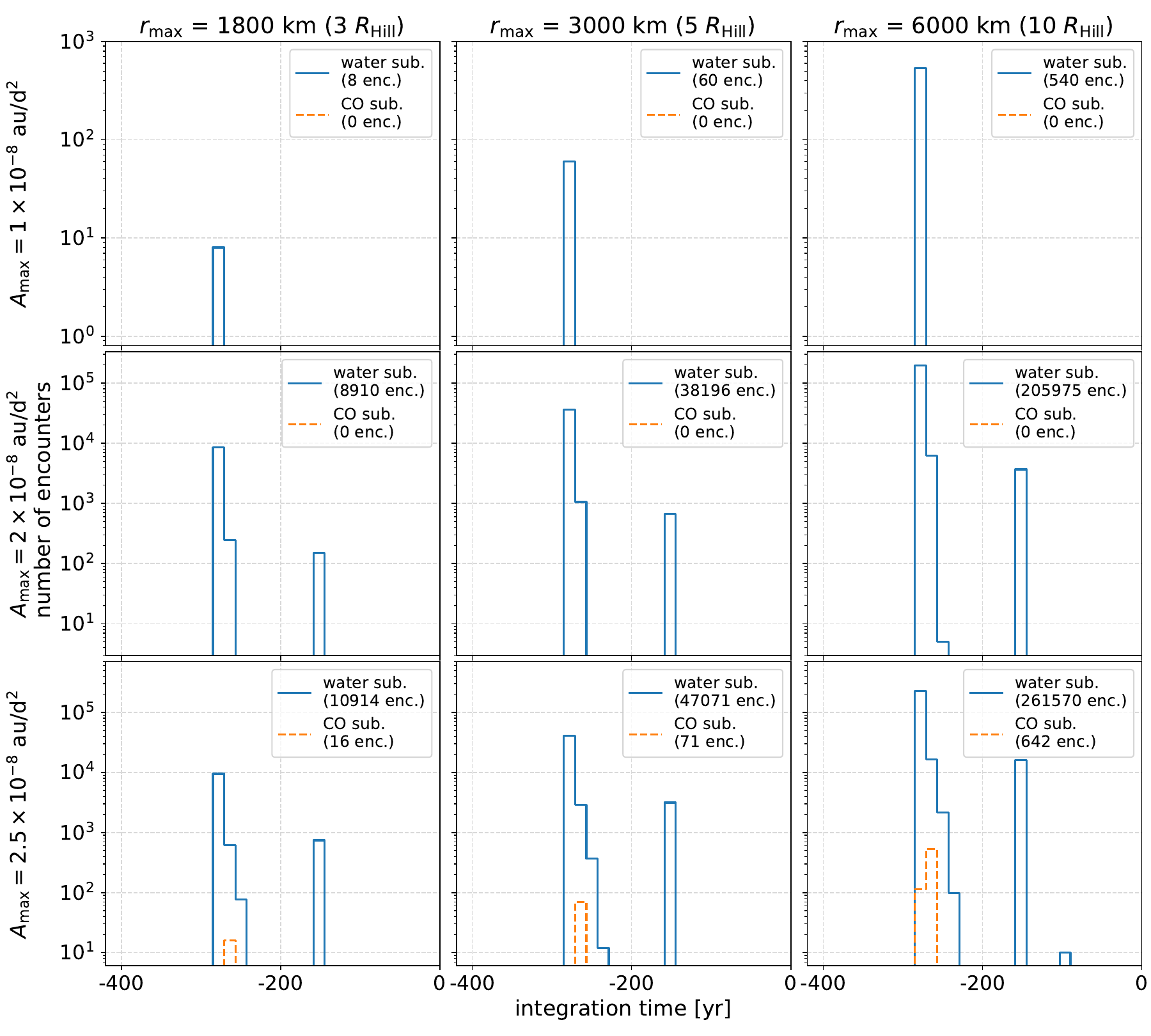}
  \caption{Histogram of close clone encounters for the water sublimation model (blue) and the CO sublimation model (orange). Each histogram bin represents one of the 28 integrations. The bin's location indicates the  period of activity and the width is equivalent to one full orbit of asteroid PR2 ($\sim 13.9$ years) around the Sun (see text in Section \ref{sec.post_breakup_activity} for more details). Each column represents a different maximum distance between clones for an encounter to be considered close. Each row displays results of integrations across different values of $A_{\rm{max}}$.}
  \label{fig.encounters_hist}
\end{figure}

Figure \ref{fig.encounters_hist} shows histograms of clone encounter times for the three chosen values of $A_{\rm{max}}$ and three values of the maximum distance threshold ($r_{\rm{max}}$) for a given encounter to be counted. Both the water and CO sublimation models are shown, and all the 28 aphelion models are included so that each histogram bin corresponds to the encounters within that one orbit of activity. It is clear that the water sublimation model allows for many more closer encounters despite having an overall weaker effect than the CO sublimation model (Fig. \ref{fig.gr_func}). In the case of the CO sublimation model, we recorded close clone encounters only in integrations with $A_{\rm{max}} \geq 2.5\times10^{-8}$ au/d$^2$, whereas in the case of the water sublimation model, close encounters were present for $A_{\rm{max}} \geq 10^{-8}$ au/d$^2$. Based on the number of clones that experience close encounters, a formation time at $T=-278.2_{-1.1}^{+0.2}$ years and, one orbit earlier, at $T=-265.2_{-1.1}^{+0.5}$ years are preferred. Another possible formation time is at $T = -152.8^{+0.3}_{-0.8}$ years, but it requires a higher value of $A_{\rm{max}}$ in the water sublimation model and is absent in the CO sublimation model for given values of $A_{\rm{max}}$. 

\begin{figure}[h!]
\centering
  \includegraphics[width=0.55\textwidth]{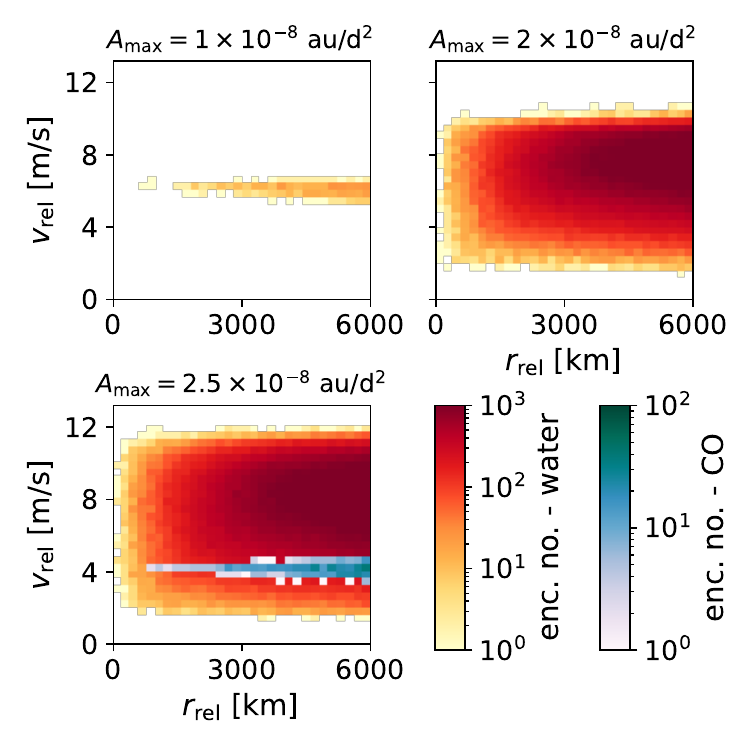}
  \caption{A heat-map visualizing number of occurrences of clone encounters at given relative distance $r_{\rm{rel}}$ and velocity $v_{\rm{rel}}$ for three values of $A_{\rm{max}}$. The yellow-to-red map represents the water sublimation model, while the blue-to-green map represents the CO sublimation model (present only in the right panel and plotted over the water sublimation model).}
  \label{fig.enc_heatmap}
\end{figure}

Figure \ref{fig.enc_heatmap} shows the distribution of relative distances ($r_{\rm{rel}}$) and velocities ($v_{\rm{rel}}$) of clones during their close encounters for the single-orbit activity model. For the water sublimation model, the lowest value of $A_{\rm{max}}$ (left panel in Fig. \ref{fig.enc_heatmap}) $v_{\rm{rel}}$ is close to 6 m/s for all 540 encounters, regardless of $r_{\rm{rel}}$. With higher values of $A_{\rm{max}}$, $v_{\rm{rel}}$ ranges from $\sim2$ m/s up to $\sim11$ m/s. This wide range of $v_{\rm{rel}}$ means that encounters with even smaller $v_{\rm{rel}}~(\approx 2$ m/s) and $r_{\rm{\rm{rel}}}~ (\approx 10^2$ km) are seen in our integrations.
In the CO sublimation model, encounters are present only for the highest value of $A_{\rm{max}}$ (right panel, Fig. \ref{fig.enc_heatmap}). These encounters are distributed around $v_{\rm{rel}} \approx 4$ m/s, similar to the distribution of encounters for the lowest $A_{\rm{max}}$ around 6 m/s in the water sublimation model.

\begin{figure}[h!]
\centering
  \includegraphics[width=1.0\textwidth]{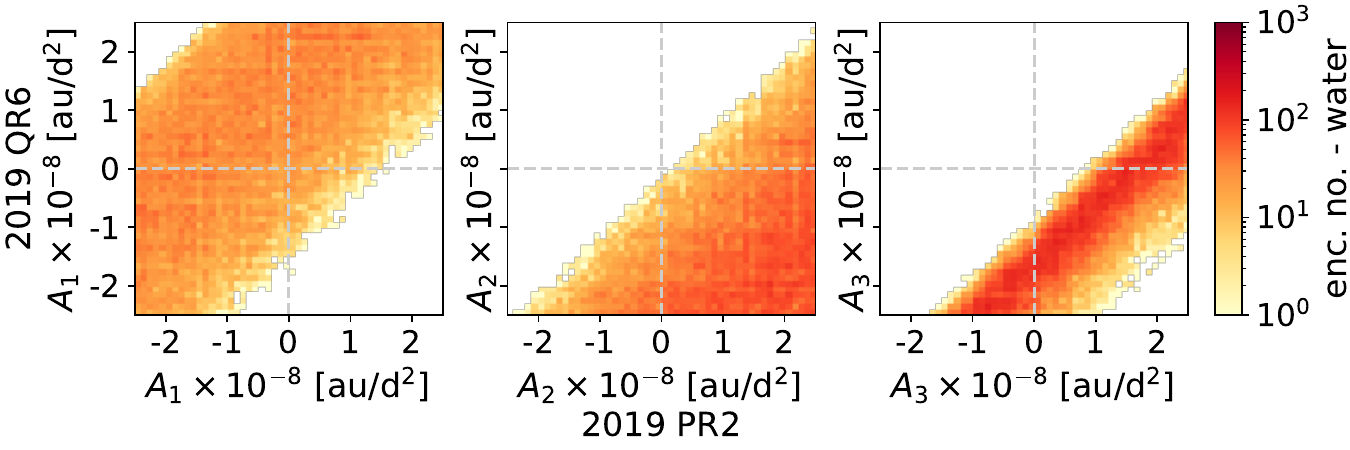}
  \caption{Distribution of $A_1, A_2, A_3$ values of clones that experienced at least one close encounter at $r_{\rm{rel}} \leq 3\,000$ km. We chose as an example an integration with the water sublimation model and with $A_{\rm{max}} = 2.5 \times 10^{-8}$ au/d$^2$ due to the high number of recorded encounters. $A_1, A_2, A_3$ values assigned to 2019~QR6 clones are displayed on the vertical axis, and values assigned to 2019~PR2 clones are displayed on the horizontal axis. Preferences for some orientations of the acceleration vectors are apparent, especially in the right panel representing acceleration in the direction perpendicular to the orbital plane.}
  \label{fig.A123_corr_heatmap}
\end{figure}

To further analyse the non-gravitational forces required for getting close encounters in this model, we checked for correlations between $A_1, A_2, A_3$ values between clones that experienced at least one encounter. Figure \ref{fig.A123_corr_heatmap} shows that despite the uniform coverage of possible values of $A_1, A_2, A_3$ within the prescribed limits (in this example $A_{\rm{max}} = 2.5 \times 10^{-8}$ au/d$^2$) in the clone generations, there are preferred certain orientations of the acceleration vector. The strongest preference is for the acceleration vector in the direction perpendicular to the orbital plane ($A_3$, the right panel in Fig. \ref{fig.A123_corr_heatmap}), where the majority of $A_3$ values lie within a $10^{-8}$ au/d$^2$ wide strip that notably does not include the point $[A_3(\mbox{PR2}), A_3(\mbox{QR6})] = [0, 0]$. In other words, we recorded close encounters with zero or close to zero values of $A_1$ and $A_2$ for both bodies, but a non-zero value of $A_3$ for at least one of the two asteroids was always required. This explains why the inclusion of the Yarkovsky effect (represented by an along track acceleration --$A_2$) in our previous integrations in Section \ref{sec.yarkovsky_effect} did not produce close encounters and thus could not explain the current orbits of this asteroid pair. Even the maximum possible strength of the Yarkovsky effect would not be sufficient in bringing PR2 and QR6 close together. For example, for the 500 meter asteroid (101955) Bennu (comparable in size with QR6), the modeled value of A$_2$ is $\sim-4.6 \times 10^{-14}$ au/d$^2$ \citep[e.g.,][]{Deo17, Farnocchia21}, which is weaker by six orders of magnitude than $A_i$ values required in our one-orbit activity integrations.

\begin{figure}[h!]
\centering
  \includegraphics[width=1.0\textwidth]{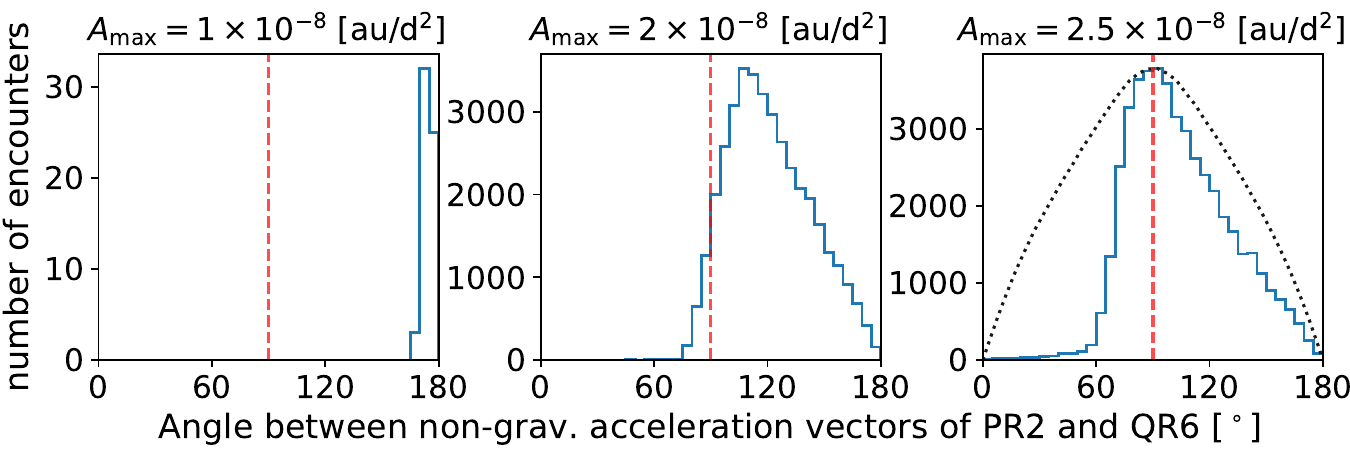}
  \caption{Distribution of angles between vectors $(A_1, A_2, A_3)$ of  PR2 and QR6 clones that experienced mutual encounters (with $r_{\rm{rel}} \leq 3\,000$ km) for the water sublimation model (blue solid line). The red dashed vertical line indicates the peak location if the $(A_1, A_2, A_3)$ vectors were random and the black dotted line in the \textit{right} panel shows the distribution of angles for randomly chosen $(A_1, A_2, A_3)$ vectors scaled in such way that its peak value corresponds to the peak value of the observed distribution. The width of each bin is $5^{\circ}$.}
  \label{fig.A_vector_angle}
\end{figure}

We calculated the angle between the vectors $(A_1, A_2, A_3)$ of PR2 and QR6 clones that experienced mutual encounters and their distribution is shown in Figure \ref{fig.A_vector_angle}. It is notable that for the lowest $A_{\rm{max}}$ value ($10^{-8}$ au/d$^2$; the left panel in Fig. \ref{fig.A_vector_angle}) the two vectors needed to be pointing in very nearly opposite directions in order that the clones could experience an encounter. For higher values of $A_{\rm{max}}$, the need for opposite direction vectors is not as profound (due to the potentially larger sizes of the vectors), but a strong preference for obtuse angle between $(A_1, A_2, A_3)$ of PR2 and QR6 clones is still evident.

\begin{figure}[h!]
\centering
  \includegraphics[width=1.0\textwidth]{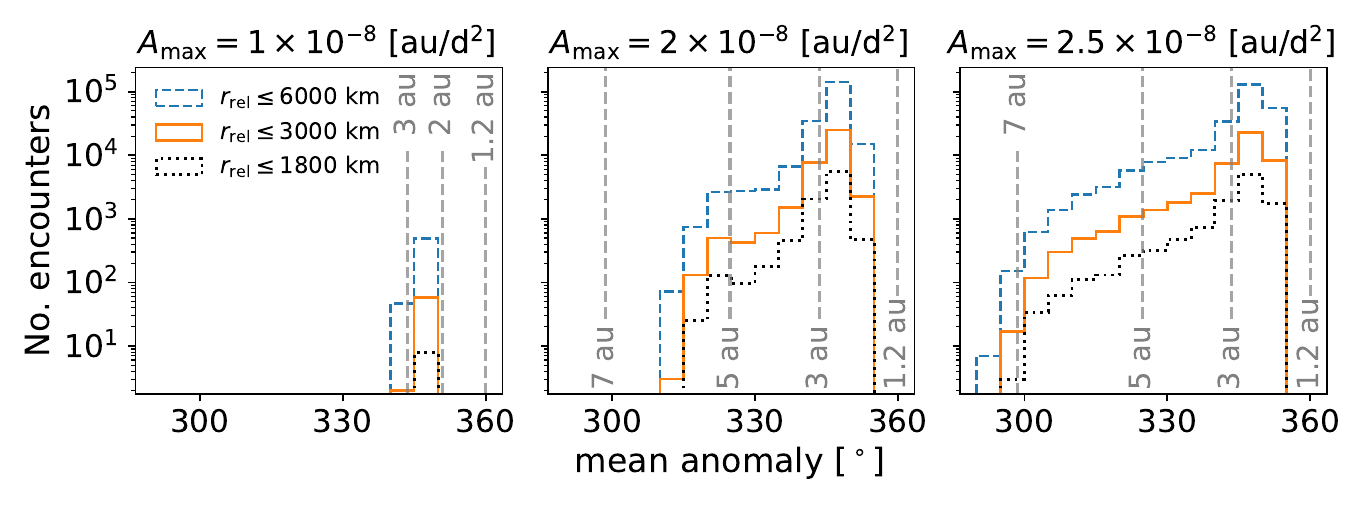}
  \caption{Distributions of mean anomalies of PR2 clones at the moment of a close encounter for three values of $A_{\rm{max}}$ in the water sublimation model. Due to the  high orbital similarity of PR2 and QR6, the mean anomaly distributions of the QR6 clones are almost identical to the one shown here. All encounters happened before the perihelion passage.}
  \label{fig.enc_loc}
\end{figure}

Due to the precisely determined initial orbits of both asteroids and the relatively short integration time needed, we were able to narrow down the probable separation location of these two asteroids. Figure \ref{fig.enc_loc} shows that the separation event happened before perihelion passage. The integration with the weakest non-gravitational acceleration ($A_{\rm{max}} = 10^{-8}$ au/d$^2$, left panel in Fig. \ref{fig.enc_loc}) suggests the separation of the two asteroids happened at mean anomaly between $340^\circ$ and $350^\circ$, corresponding to heliocentric distances between $\sim3.4$ and $\sim2.2$ au before perihelion. For the two higher values of $A_{\rm{max}}$, separation events occurring at larger heliocentric distances were possible.

In this section, we showed that the current orbits of PR2 and QR6 can be explained by a gentle separation (possibly by rotational fission) of their parent body that was followed by a short-term (up to a maximum of one full orbit) of comet-like activity that perturbed their orbits. Unlike in the model with continuous activity presented in the previous Section, the water sublimation model produced significantly more clone encounters than the CO sublimation model for the short-lived one-orbit activity. Both sublimation models suggest that the separation of the two asteroids occurred around $T\sim-270$ yr (Fig. \ref{fig.encounters_hist}) and that the separation happened before a perihelion passage (Fig. \ref{fig.enc_loc}). We found that at least one of the two asteroids was accelerated in the direction normal to its orbital plane (the right panel in Fig. \ref{fig.A123_corr_heatmap}).

\subsection{Future orbital evolution}
\label{sec.future_orb_evol}
We integrated clones of PR2 and QR6 into the future to evaluate their further orbital evolution. For this, we assumed only gravitational forces act on the two asteroids, i.e., neglected possible Yarkovsky effect. Orbital clones of both asteroids remain compact (deterministic) until an encounter with Jupiter at $\rm{T} \sim+747$ yr at a distance $\sim6.9\times10^7$ km from Jupiter. The clones of QR6 started to disperse rapidly after this encounter. A subsequent encounter with Jupiter at $\rm{T} \sim+877$ yr accelerated the dispersion of PR2's clones. At times around $+3\,000$ yr, first clones with hyperbolic orbits started to appear, suggesting integration instability.

\section{Formation and Evolution}
\label{sec.formation}

\subsection{Primary Rotational State}
\label{prim_rot_discus}
Rotational fission caused by YORP spin-up is thought to be the predominant mechanism for asteroid pair formation \citep{Pravec19}. From theoretical considerations of the free energy and angular momentum conservation, the size ratio of a YORP-fissioned asteroid pair is related to the rotation period of the larger, primary member of the pair. Following that theory, the size ratio of PR2 and QR6 as estimated from the difference of their absolute magnitudes $\Delta H = 1.2$ \citep[see Table \ref{tab.orbits}; an uncertainty of the absolute magnitude difference is estimated to be $\pm 0.3$~mag, see][]{Pravec19}, suggests that the period of PR2 should be greater than about 4~h \citep[see Fig.~34 in][]{Pravec19}. The possible periods of PR2 obtained from our observations in Section~\ref{sec.rotations} are in the theoretically predicted range, so the data are consistent with the theory.

The possibility of asteroid PR2 being in a non-principal axis rotation state (so-called tumbling), suggested by the poor fits in some parts of its lightcurve  (see Section~\ref{sec.rotations}), is interesting.  As discussed in Section~6 of \citep{Pravec19}, it is remarkable that all of the paired asteroids (which are older by more than one order of magnitude than PR2--QR6) for which they obtained sufficient observational data showed just one-period rotational lightcurves, i.e., there was present no apparent tumbling in any of them.  As they estimated the median damping time of a non-principal axis rotation of the paired asteroid primaries to be about $5 \times 10^5$\,yr, i.e., greater than the ages of many of the observed asteroid pairs, the fact that they saw no tumbling in any of them gives an important constraint to the theories of asteroid pair formation and evolution.

If the suspected tumbling of PR2 is confirmed, it will be a unique finding that could shed more light on the formation of the PR2--QR6 asteroid pair.  For instance, we could speculate that the rotation of PR2 might be more excited ---and thus producing an observable tumbling signal in its lightcurve; see the note on the minimum detectable rotational axis misalignment angle in Section~6 of \citep{Pravec19}--- than the primaries of ordinary main-belt asteroid pairs, possibly by the cometary activity that apparently followed the formation of this asteroid pair.  A confirmation of the suspected tumbling of PR2 with thorough and high-accuracy observations around its future perihelion passage is highly needed.

\subsection{Composition and Surface Properties}

We showed in Section \ref{sec.observations} that the photometric colors of both PR2 and QR6 are relatively close to a $D-$type taxonomic class. This is an unusual classification for asteroids in general and is only seen in $<10\%$ of observed NEAs \citep{Devogele19}. The $D-$type taxonomic classification is more common amongst comets, but only 115 near-Earth comets are known\footnote{According to the \textit{Center for Near-Earth Object Studies} discovery statistics available at \texttt{https://cneos.jpl.nasa.gov/stats/totals.html}.} (as of August 2021) and finding two unrelated comets with such similar orbits and properties is extremely unlikely. The similar spectral properties and rarity of this taxonomic class, particularly amongst NEAs, strongly support a common origin for this pair.

These spectral properties are indicative of an origin  outside of the Main Belt \citep[e.g.,][]{Hasegawa21}. As these object's colors are intermediate to that of Red Group Trojans and the reddest of Centaurs (Fig. \ref{fig.colors}), it is unclear whether PR2 and QR6 originally (prior to pair formation) came from the Jovian Trojan clouds or the Kuiper Belt. Either possibility raises interesting questions about color changes related to these object's transition into NEO orbits. Would a Jovian Trojan become redder under increased heating and irradiation as an NEO? Would an object from the Kuiper Belt become bluer under these conditions? NASA's Lucy mission to several Jovian Trojans \citep{Levison16} may help provide insights relevant to these questions.

Despite the color similarity, we noted that PR2 has a steeper spectral slope than QR6, primarily due to a difference in $r-i$ color. We suggest causes of this offset may include differences in observing geometry, object size, composition, surface grain size, and/or evolutionary effects. It remains unclear how any of these causes would affect just $r-i$ color and not the other two colors that were measured. High quality spectral observations are encouraged at the time of these object's next perihelion passage.

Since these objects are on such similar orbits, even down to a difference in mean anomaly of $<0.01^\circ$ (Table \ref{tab.orbits}), the observing geometries (e.g. phase angle) on a given night of observation were nearly identical. As such, we do not expect the measured color difference to be attributable to geometry effects such as phase reddening \citep[e.g.,][]{Millis76,Lumme81,Sanchez12}.

Color variations, comparable in magnitude to the difference between PR2 and QR6 ($\sim5\%/0.1\mu$m), have been detected for $D$-type asteroids as a function of object size \citep[e.g.,][]{Jewitt90,Fitzsimmons94,Gartrelle21}. This trend shows that smaller objects are generally redder. The origin of this effect is unknown, but could be attributed to differences in strength for objects of different compositions, e.g. redder compositions may be structurally weaker and fragment under collisions more easily. However, this trend applies to objects that are 10's to 100's of km in size, orders of magnitude larger than PR2 and QR6, and is specific to objects in the Jovian Trojan population. Given that the larger member of our NEO pair is redder, the opposite of that seen for Trojan $D$-types, and that size-color trends for $D$-types have not yet been seen for small NEOs, we think it is unlikely that the 2:1 size ratio of PR2 to QR6 contributes to their color difference.

From a compositional perspective, iron-rich silicates and tholin-like organics may be responsible for reddening surfaces in the outer Solar System \citep[e.g.,][]{Sharkey19,Hasegawa21}. However, given the complexity of pair formation from a rubble-pile precursor \citep{Walsh08,Scheeres07}, it is unlikely that any separation event could have produced components with significantly different compositions. 

Subsequent evolution of the pair after separation could have involved a temporary phase as an unstable binary with ongoing disruption of the secondary due to tidal interactions with the primary \citep{Jacobson11}. These secondary disruption events imply the possibility of a different evolutionary history for QR6 relative to that of PR2. Such post-separation differential evolution was suggested by \citet{Polishook14} as a way to explain slope differences between components in two S-complex asteroid pairs in the Main Belt. However, without fully understanding details of the pair formation process it is hard to know whether this is a viable explanation.

We do note that if the pair formation process produced two objects with similar surface properties, then the present day color difference can not easily be attributed to the effects of space weathering (or any of the other previous suggestions). Even though we do not have a clear understanding of how space weathering affects $D$-type asteroids or outer solar system bodies \citep[e.g.,][]{Lantz18}, as an asteroid pair, PR2 and QR6 have experienced similar levels of exposure to irradiation, seasonal thermal cycles, and particle bombardment. 

Given the preceding discussion, it is plausible, if not likely, that PR2 and QR6 emerged several hundred years ago as a separated pair with different surface characteristics. A particularly intriguing possibility that could explain our results is that these two objects may have different surface grain size distributions. Grain size is known to affect spectral slope. For example, \citet{Cloutis18} suggest that for dark, primitive asteroids smaller grain sizes cause redder spectral slopes. This would imply that PR2 has finer surface grains than QR6. This would not be too surprising for bodies with mature surface regolith ---in general larger objects have smaller surface grains--- but the timescales of regolith formation are thought to be orders of magnitude longer than the age of this pair \citep{Shkuratov01,Willman10}. In other words, even if the size difference between these bodies was enough to result in the formation of regoliths with different grain size distributions, there simply has not been enough time for this maturation process to reach a steady state. Therefore, any difference in surface grains would have been inherited at the time of the pair formation process.

This scenario of different inherited grain sizes is particularly appealing because it can be tested, both directly and indirectly. Radiometry at thermal infrared wavelengths can be used to measure the thermal inertia of a planetary surface, which in turn is directly related to surface grain size \citep{Gundlach13}. Targeted observations in 2033, for example with the James Webb Space Telescope, during the next perigee passage of these objects could directly probe for differences in thermal inertia. Indirect tests may come from NASAs upcoming DART \citep{Rivkin21} and JANUS \citep{Scheeres21} missions, both of which will be visiting binary asteroids. These missions may constrain any grain size differences between the primaries and satellites, which will provide important clues into the detailed physics of binary and asteroid pair formation and evolution.

\subsection{Dynamical history}

From our backwards orbital integrations presented in Section \ref{sec.orbital_integ}, we showed that in the pure gravity regime, PR2 and QR6 do not get sufficiently close to each other before a close encounter with Jupiter at $T\sim -406$~yr. The Jovian encounter perturbs PR2 and QR6 orbits differently making any close clone encounters prior to that Jovian flyby very unlikely. This led us to suspect that either PR2 and/or QR6  experienced  significant non-gravitational accelerations sometime between the present and the Jovian encounter, and that the cause of the acceleration might be linked to their formation as an unbound pair.

We demonstrated that the acceleration caused by the Yarkovsky effect (represented by an along-track acceleration) is not effective and does not bring the PR2 and QR6 clones much closer together. Only small differences in encounter distances between clones were observed for geometric albedos (linked to the Yarkovsky effect strength) ranging from 0.02 up to 0.5.

With the use of the Marsden model describing acceleration induced by cometary activity, we obtained close encounters between clones of PR2 and QR6. We tested a model with continuous activity and a model with activity lasting for up to one complete orbit. Both models suggest a similar separation time around $T \approx -300 $ years (with other solutions possible, but statistically less probable). Interestingly, the model including cometary activity during each orbit produced more close encounters for the CO sublimation model, whereas in the short-term activity model, more encounters were detected in the water sublimation model. We note that both models (continuous activity and one-orbit activity) can be considered as extreme cases and the real separation event scenario may be somewhere in between (e.g., a gradually decreasing activity over several orbits). Therefore, it is not possible to reveal the chemical composition of PR2 or QR6 by models used in our orbital integrations.

The continuous activity model allowed for encounters with relative velocities comparable with the escape velocity from the larger component PR2, which is in an agreement with the rotational fission theory, suggesting that PR2 and QR6 formed a temporary stable proto-binary system for a short time \citep{Scheeres07}. The temporary orbit of the secondary QR6 around the primary PR2 was within Hill's radius of the primary. Therefore, we looked for mutual clone encounters within a few Hill's radii \citep[following][]{Pravec19, Fatka20}. In the one-orbit activity model, the relative velocities during encounters were larger than the escape velocity from PR2 by a factor of few. This might be due to the simplification of the acceleration induced by the sublimation activity in our model. 

Both our comet-like activity models indicate non-zero values of $A_3$ (out-of-the-orbital-plane acceleration) for one or both bodies (Fig. \ref{fig.A123_corr_heatmap} and in Appendix \ref{fig.encounters_hist_350_appendix}). The acceleration in the direction out of the orbital plane might have several causes, such as that the active region (producing jets) was not located on the equator of the active body or that the volatile substances (propellant) are distributed asymmetrically on its surface. 

\section{Summary}
\label{sec.summary}

The preceding sections have demonstrated the following for the NEO pair 2019~PR2 -- 2019~QR6:
\begin{itemize}
\item We recovered archival detections of both objects dating back to 2005. Their refined orbits are deterministic for hundreds of years on either side of the current epoch, from $-406$~yr to $+747$~yr, at which times close encounters with Jupiter cause major perturbations to the orbits.

\item They are spectrally close to $D$-type asteroids with spectral slopes that are steeper than typical Trojan asteroids but less red than the reddest objects in the Centaur and Kuiper Belt populations. If the parent body of PR2 and QR6 originated in one of these outer Solar System reservoirs, then this pair provides insight on the color affects of increased heating and irradiation experienced in an NEO orbit.

\item PR2 has a steeper spectral slope than QR6 due to a redder SDSS $r-i$ color. This color difference is most likely due to differences in surface properties inherited from the pair formation process. One possible difference could be these object's surface grain size distributions. Future confirmation of this would provide important insight into the details of the pair formation process.

\item The most plausible/candidate rotation periods of PR2 and QR6 are 9.9~h and 4.1~h respectively, but neither period is strongly constrained.

\item An apparent offset of the overlapping data in the composite lightcurve of PR2 might be caused by a non-principal axis (NPA) rotation.  If confirmed with future thorough and high-quality observations, it will be a unique finding as a NPA rotation has not been detected in ordinary main-belt asteroid pairs yet.

\item PR2 does not display any sign of ongoing cometary activity.

\item Mere gravitational attraction in backward orbital integrations fails to bring PR2 and QR6 close to each other before the relatively close encounter with Jupiter at $T\sim -406$ years, which puts PR2 and QR6 into very different orbits before it.

\item The Yarkovsky effect is not effective to get PR2 and QR6 close enough to each other  between $T = -406$ years and the present.

\item Implementation of a Marsden-style model of non-gravitational acceleration due to cometary activity allowed for close encounters to appear in both tested models (continuous activity and one-orbit activity models). Both models suggest a separation time around $T\sim-300$ years. Interestingly, the continuous activity model produced significantly more encounters for the CO sublimation model, whereas the one-orbit activity model produced more encounters for the water sublimation model. With the real orbital and activity history being probably in between the two tested (extreme case) models, we cannot constrain the chemical composition of PR2 and QR6 by the dynamical simulations.

\item The combination of $(i)$ eccentric orbits ($e\approx0.8$) and Tisserand's invariant $T_{Jup} = 2.15$, $(ii)$ taxonomy close to $D$-type classification with steep red-sloped spectra, and $(iii)$ necessity of comet-like activity for clone encounters to appear in our backward orbital integrations, raises the question whether PR2 and QR6 are fragments of a parent asteroid or fragments of a dormant comet.

\end{itemize}

{\bf Data availability}\\
The data underlying this article will be shared on reasonable request to the corresponding author.\\

{\bf Acknowledgements}\\
We are grateful to the reviewer David Vokrouhlick\'y for his useful comments and suggestions that let us improve this paper.
PF, PP and PK were supported by the Grant Agency of the Czech Republic, Grant 20-04431S. NM, AG and MD acknowledge support from a NASA Near Earth Objects Observations grant NNX17AH06G, awarded to the Mission Accessible Near-Earth Object Survey (MANOS). MD and NM acknowledge support from a NASA Near Earth Objects Observations grant 80NSSC21K0045 awarded to the Second lunation NEO follow-up. NM and JK acknowledge support from NASA's Hayabusa2 participating scientist program (grant NNX16AK68G). Based on observations with the NASA/ESA Hubble Space Telescope obtained [from the Data Archive] at the Space Telescope Science Institute, which is operated by the Association of Universities for Research in Astronomy, Incorporated, under NASA contract NAS5-26555. Support for Program number HST-GO-14133.002-A was provided through a grant from the STScI under NASA contract NAS5-26555.

\clearpage

\appendix
\section{}

\afterpage{%
\clearpage
\begin{landscape}
\begin{table}[hbt]
\footnotesize
\caption{Orbital elements with non-gravitational parameters $A_1, A_2, A_3$ fitted by the \texttt{Find Orb} software for the water and CO sublimation models.
Vital stats for 2019 PR2 and 2019 QR6: absolute magnitude (H), and orbital elements semi-major axis ($a$), eccentricity ($e$), inclination ($i$), longitude of ascending node ($\Omega$), argument of perihelion ($\omega$), and mean anomaly ($M$) from our updated orbit solution including pre-covered astrometry data from 2005 (see Section \ref{sec.observations}) . Orbital elements are derived for epoch JDT=2458800.5 (2019-Nov-13.0 TT) and precision extends to the last significant digit in each case.}

\begin{center}
\begin{tabular}{llllllllll}
\label{tab.orbits_A123_appendix}

Object	    		& $a$ [au]   & $e$ [~]     & $i$ [$^\circ$] & $\Omega$ [$^\circ$]  & $\omega$ [$^\circ$]  & $M$ [$^\circ$]   & $A_1\times10^{-10}$ [au/d$^2$] &  $A_2\times10^{-10}$ [au/d$^2$]  & $A_3\times10^{-10}$ [au/d$^2$] \\

\hline
\multicolumn{10}{l}{\textit{water sublimation model}} \\
2019 PR2    & 5.77195 & 0.797683 & 10.989902 & 349.03451 & 57.09581 & 1.31518 & $-1.7 \pm 2.9$ & $-0.1 \pm 2.1$ & $1.7 \pm 0.9$\\
2019 QR6    & 5.77264 & 0.797720 & 10.971805 & 348.99835 & 57.12938 & 1.30110 & $-3.1 \pm 3.5$ & $-0.3 \pm 2.7$ & $1.0 \pm 1.0$\\
\multicolumn{10}{l}{\textit{CO sublimation model}} \\
2019 PR2    & 5.77201 & 0.797685 & 10.989914 & 349.03453 & 57.09578 & 1.31516 & $-1.7 \pm 2.7$ & $1.5 \pm 0.9$ &  $2.0 \pm 1.0$\\
2019 QR6    & 5.77268 & 0.797722 & 10.971809 & 348.99836 & 57.12935 & 1.30109 & $-1.6 \pm 3.0$ & $1.3 \pm 1.1$ &  $0.9 \pm 1.0$\\
\hline
\end{tabular}
\end{center}
\end{table}
\end{landscape}
\clearpage
}


\begin{figure}[h!]
\centering
  \includegraphics[width=.80\textwidth]{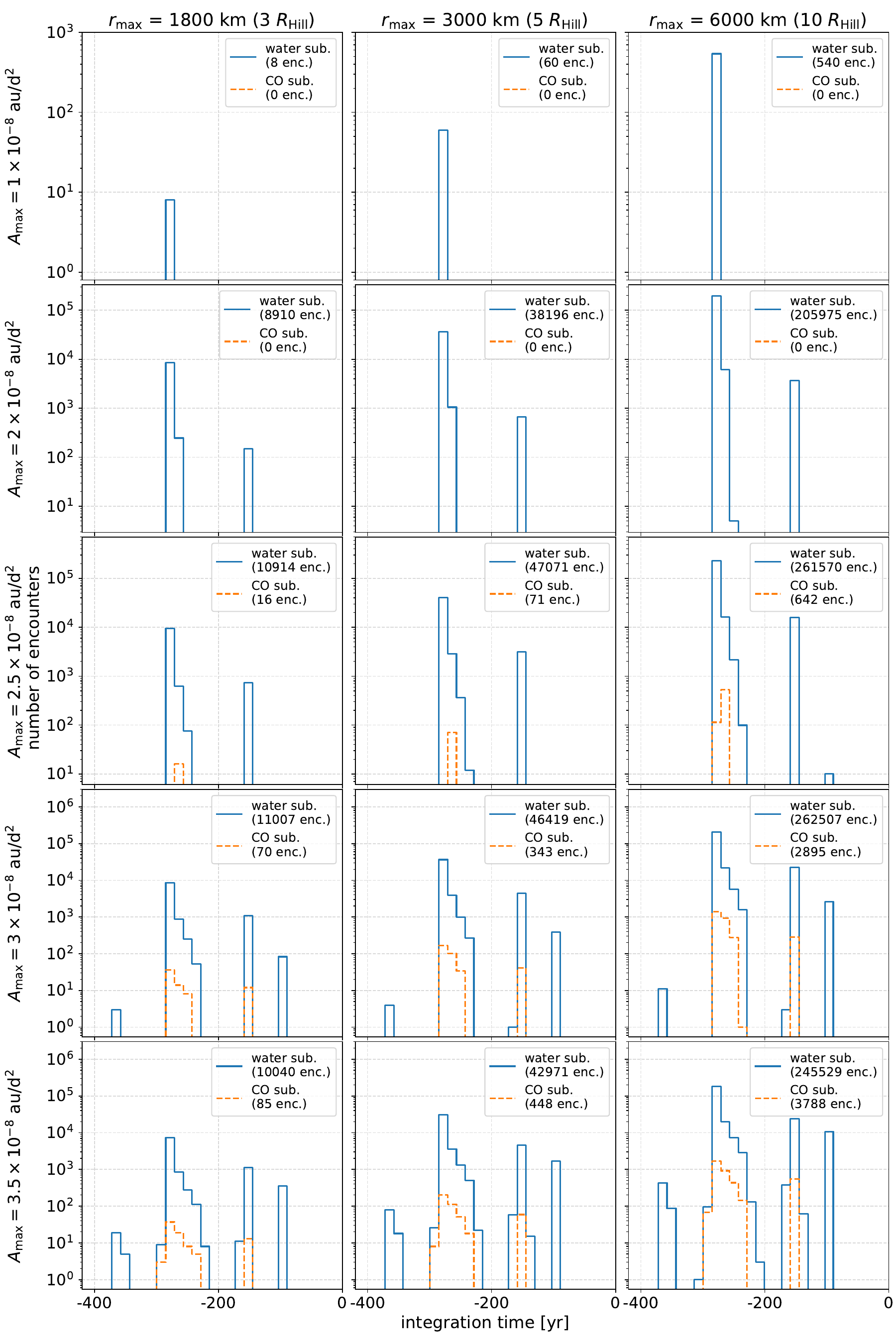}
  \caption{Same as Figure \ref{fig.encounters_hist} in Section \ref{sec.one-orbit-model}, but results of integrations for higher values $A_{\rm{max}}$ are included.}
  \label{fig.encounters_hist_350_appendix}
\end{figure}

\clearpage




\bibliographystyle{apalike} 

\bibliography{references.bib}
\biboptions{round, authoryear}



\end{document}